\documentclass{jocg}

\usepackage{hyperref}
\hypersetup{colorlinks=true, linkcolor=Blue, filecolor=RawSienna, citecolor=Purple, urlcolor=OliveGreen}
\usepackage{amssymb}
\usepackage{amsmath}
\usepackage{amsthm}
\usepackage{amsfonts}
\usepackage{amscd}
\usepackage{graphicx}
\usepackage[dvipsnames]{xcolor}
\usepackage[left]{lineno}

\usepackage{fullpage}
\usepackage{float}

\usepackage{graphics,amssymb}
\usepackage{latexsym}

\usepackage{url}
\usepackage{appendix}
\usepackage{adjustbox}

\usepackage[nameinlink]{cleveref}
\usepackage[hyperpageref]{backref}
  \renewcommand*{\backref}[1]{}  
\renewcommand*{\backrefalt}[4]{
	\ifcase #1 
	No cited.
	\or
	(page #2.)
	\else
	(pages #2.)
	\fi}

\makeatletter
\let\c@figure\c@table
\makeatother 

\newcommand{\seqnum}[1]{\href{https://oeis.org/#1}{\rm \underline{#1}}}
\newcommand{\la}{\ensuremath{\langle }}
\newcommand{\ra}{\ensuremath{\rangle}}
\newcommand{\Lr}[1]{\ensuremath{\big\langle{#1}\big\rangle}}
\newcommand{\LR}[1]{\ensuremath{\Big\langle{#1}\Big\rangle}}
\newcommand{\lr}[1]{\ensuremath{\langle{#1}\rangle}}
\newcommand{\m}[1]{\ensuremath{\mathrm{#1}}}
\newcommand{\N}{\ensuremath{\mathbb{N}}}
\newcommand{\R}{\ensuremath{\mathbb{R}}}
\newcommand{\Rev}{\ensuremath{\mathcal{R}}}
\newcommand{\s}{\ensuremath{ \sigma}} 
\newcommand{\Z}{\ensuremath{\mathbb{Z}}} 
\newcommand\mapsfrom{\mathrel{\reflectbox{\ensuremath{\mapsto}}}}
\DeclareMathOperator{\sgn}{sgn} 
\DeclareMathOperator{\ng1}{neg} 
\DeclareMathOperator{\inv}{inv} 

\crefformat{figure}{#2fig.~#1#3}
\Crefformat{figure}{#2Figure~#1#3}
\crefformat{definition}{#2def.~#1#3}
\Crefformat{definition}{#2Definition~#1#3}
\crefformat{section}{#2sec.~#1#3}
\Crefformat{section}{#2Section~#1#3}

\begin{document}

\theoremstyle{definition}
\newtheorem{theorem}{Theorem}
\newtheorem{lemma}[theorem]{Lemma}

\theoremstyle{plain}
\newtheorem{remark}[theorem]{Remark}
\newtheorem{definition}[theorem]{Definition}
\newtheorem{example}[theorem]{Example}
\newtheorem{observation}[theorem]{Observation}
\newtheorem{corrolary}[theorem]{Corrolary}
\newtheorem{conjecture}[theorem]{Conjecture}

\theoremstyle{remark}

\begin{center}
	{\LARGE\bf 
	Fractal Images as Number Sequences I}\\
	\vskip 5mm	{\Large An Introduction}
	\vskip 1cm
	\large
			Arie Bos
			
			Mathematics and Computer Science

			Technological University Eindhoven 
			
			PO Box 513, 5600 MB, Eindhoven
			
			The Netherlands
			
			\href{mailto:a.bos.1@tue.nl}{\tt a.bos.1@tue.nl} 
\end{center}

\vskip 0.2in
\begin{abstract} 
	In this article, we considered a fractal image as a fractal curve, that is, as a walk on a grid in Euclidean space $\R^d$.
	We placed integers on the generating vectors of a grid, such that opposite directions have opposite numbers. This numbering system converts a curve on that grid into a sequence of integers, corresponding with the curve's edges. The corresponding sequence contains the same fractal structure, i.e., an approximant of the curve corresponds to that of the sequence.	
	We introduced a normalized sequence which is unique for a curve.
	The morphisms of the grid generators were translated into signed permutations on the alphabet of all the numbers used.	
	By ordering the fractal sequences, we obtained an encyclopedia of fractals.
	A variety of examples and images enriched the text.
\end{abstract}

\section{Introduction}
\begin{figure}[H]
	\centering
	\includegraphics[scale=1.1]{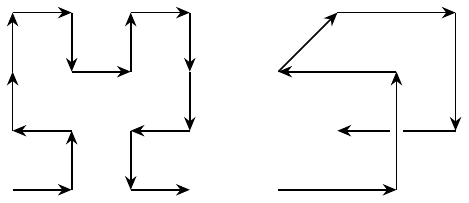}
	\caption{\small Hilbert's second approximant and the third approximant of the Gray curve.} 
	\label{fig:introd1}
\end{figure}

Assuming we have both an integer sequence $\lr{1,2,-1,2,2,1,-2,1,2,1,-2,-2,-1,-2,1}$ and a drawing rule that states for $1$ you draw one unit to the right, for $-1$ one unit to the left, for $2$ draw one unit up, and for $-2$ one unit down. The drawing would then be as shown in the left picture in \cref{fig:introd1}.
Alternatively, we provide the sequence $\lr{1,2,-1,3,1,-2,1}$, plus the rule that $3$ is one unit backward, then the drawing is as shown in the right picture in \cref{fig:introd1}.
The main objective of this article is:~\emph{to describe a system in which we can represent fractal curves in all dimensions as integer sequences.}

\vspace{0.5\baselineskip}
In \cref{sec:axiomdscrpt} we give an axiomatic description of the tools required to handle the two objects in the paper, sequences and fractals.

\Cref{sub:Asmn} is dedicated to defining the basics surrounding sequences, that is the alphabet, including its inverses, morphisms, and the property of being normalized. Additionally, we also define some unique morphisms because of their vital role. 
The second subsection we dedicate to a peculiar morphism sub-type, represented by  signed permutations on the alphabet.
\Cref{sub:gdi} defines what we need for fractals, i.e., grids, directions, and isometries.
The following sub-section describes the other object from the title, i.e., fractals and their approximations.
In \cref{sub:ventrflgs} a geometric tool gives us the possibility to indicate the substitution of that edge.  
Finally, in \cref{sub:reprfrun} we propose a concise way to display a fractal and all its characteristics.

The section \nameref{sec:exmpls} is devoted to some divergent examples where our method is applied.
The last section discusses conclusions and considerations.

In the appendices, \cref{sc:Cayley} is for an overview of the dihedral group of transformations of the square grid, whereas the second one, \nameref{sc:NrmFrEnc}, is for a setup and start of an encyclopedia of fractals.

\begin{remark}
This article uses one numbering for (sub)sections and another for figures, definitions, theorems, remarks, lemmas, observations, and examples.
However, there is no rule without exception: occasionally, we will use a framed box of text that we think is important enough to call this block an \emph{Intermezzo}.
\end{remark}

\section{Axiomatic description}\label{sec:axiomdscrpt}

\subsection{Alphabet, sequences, morphism, and normalized}\label{sub:Asmn}
In this subsection we sketch four definitions for: alphabet, sequence, normalized, and morphism, respectively, to build our environment of sequences.
Notice we use a less standard definition for an alphabet, including the inverse elements, and using natural numbers plus their negatives. 
Consequently, \emph{integer sequences}, together with concatenation as operand, directly constitute a free group over this alphabet. 
In the morphisms on that group we distinguish three particular kinds, being reverse, inverse, and substitution, which appear to be helpful in our approach.
This subject's thorough and more general treatment is provided in \cite{allshal}.
\begin{definition}
 An \textbf{alphabet} of \emph{order} $n$ is
 $\Sigma_n=\{k|\, k\in \Z; 0<|k|\le n \}$, for some \mbox{$0<n\in \N$,} a subset of the integers, that we abbreviate to $\Sigma_n=\big\{\pm 1,\pm 2,\ldots,\pm n\big\}$.
 For $n=\infty$ we have the infinite alphabet $\Sigma_\infty=\Z\setminus\{0\}$.
\end{definition}

\begin{definition}
 An \textbf{integer sequence}, also known as \emph{a sequence}\footnote{~In sequence theory, this is usually called a \emph{word}, since the alphabet is often a set of \emph{letters}. However, we only use integers.}, is a countable, \emph{ordered multiset} with elements taken from an alphabet $\Sigma_n$.
 Let $\Sigma_n^*$ constitute the set of all finite sequences, where we denote each sequence by $S=\lr{s_1,s_2,\ldots}$\footnote{~Eventually, we use $S=\Lr{s(1),s(2),\ldots}$ with $s(k)\in \Sigma_n$} with $s_k\in \Sigma_n$, and \emph{with commas and angle brackets}.
 This notation represents the empty sequence by $\lr{}=\epsilon$.
 By $|S|$ we denote the \emph{absolute} sequence $|S|=\Lr{|s_1|,|s_2|,\ldots}$, whereas the \textbf{length} of a sequence, i.e., its number of elements, we indicate by $\|S\|$.
 Let $\Sigma_n^k=\big\{S\in \Sigma_n^*;\, \|S\|=k\big\}$ for $0\le k$, be the set of sequences with length $k$, so $\Sigma_n^0=\lr{}$, then $\Sigma_n^*=\bigcup\limits_{k\ge 0}\Sigma_n^k$.	
\end{definition}
For clarity, we use commas to separate items within a sequence to avoid confusion between the different numbers.
To introduce a \emph{normalized sequence}, we define a \emph{(normalized) ordering} of the non-zero integers as follows: $1,-1,2,-2,3,-3,\ldots$. That is, a positive integer comes before its negation, the positive numbers occur in ascending order, and the negative numbers in descending order, i.e., $n\prec m  \Leftrightarrow (|n|<|m|) \vee (0<n=-m)$.
A sequence $S$ of integers we call normalized if the integers in the sequence $S$ appear to be ordered in a normalized way. More precisely:

\begin{definition}\label{def:normalized}
	For a sequence $S=\lr{s_1,s_2,\ldots}$ and for $0< k\in \Sigma_n$, let $n_k$ be the lowest index $i$ such that $s_i = k$, and $n_k=\infty$ if $k\notin S$, i.e., $s_{n_k}=k$ and for $j<n_k$ we have $s_j\not=k$. 
	This sequence is \textbf{normalized} if $|s_j|<k$ for $1\le j<n_k$ and $0< k\in \Sigma_n$.
	Therefore, for a normalized sequence, we have $n_1 = 1$, and $n_{k-1}<n_k$ for $k > 1, k\in \Sigma_n$.
\end{definition}

A \emph{finite} sequence can be normalized in two ways because its reverse can also be normalized, for instance, $\lr{1,2,-1,1}$. In this case, we prefer the lexicographically smaller of the two. Therefore, $\lr{1,-1,2,1}$ will be the minimal normalized version.
See \cref{df:seqsort} (page \pageref{df:seqsort}) for how we order sequences.

Because the set $\Sigma_n^*$ is a monoid (even a \emph{free group}, as we will see) with concatenation (of sequences) as multiplication, denoted by a \emph{comma}, and $\lr{\;}=\epsilon$ as the identity element, it is natural to consider mappings that respect this operation.

\begin{definition}\label{def:morhism}
A mapping $\phi:\Sigma_n^*\to \Sigma_n^*$ such that $\phi(S,T)=\big(\phi(S),\phi(T)\big)$, where $S,T\in \Sigma_n^*$ is a homomorphism, or \textbf{morphism} for short.
\end{definition}
\noindent However, some unique morphisms deserve a memorable name: reverse, inverse, and substitution.
\begin{definition}\label{def:reverse}
	 We define the \textbf{reverse}, which we denote by $\Rev$, for a sequence $\lr{s_1,s_2,\ldots,s_n}$, by This $\Rev\lr{s_1,s_2,\ldots,s_n}=\lr{s_n,s_{n-1},\ldots,s_1}$. $\Rev:\Sigma_n^*\to \Sigma_n^*$ is a peculiar mapping because it is an \emph{anti-homomorphism}, i.e., $\Rev(S,T)=\big(\Rev(T),\Rev(S)\big)$ for $S,T\in \Sigma_n^*$.
	Notice that $\Rev$ is only defined for sequences with finite length and that $\Rev\lr{x}=\lr{x}$ for $x\in \Sigma_n$.
\end{definition}
There is a natural embedding of $\Sigma_n$ into $\Sigma_n^*$ by the injection $x\mapsto \lr{x}$; therefore, we identify $\Sigma_n$ with $\Sigma_n^1$.
With $\alpha:\Sigma_n\to \Sigma_n^*$ a morphism, there is a natural extension to $\alpha^*:\Sigma_n^*\to \Sigma_n^*$ by $\alpha^*\lr{s_1,s_2,\ldots,s_k}=\lr{\alpha(s_1),\alpha(s_2),\ldots,\alpha(s_k)}$. However, we will use $\alpha$ instead of $\alpha^*$.

A morphism $\s$ on $\Sigma_n^*$ with $\|\s(x)\|=1$ or $x\in\Sigma_n$ we call a \emph{coding}.
	Defined on $\Sigma_n$ only, it extends naturally to a bijection on $\Sigma_n^*$.
For a coding $\alpha:\Sigma_n\to \Sigma_n$, its extension $\alpha:\Sigma_n^*\to \Sigma_n^*$ is \emph{length-preserving}, i.e., $\|\alpha(S)\|=\|S\|$ for $S\in \Sigma_n^*$.
Examples of trivial codings on $\Sigma_n$ are the identity we denote by $\iota$ and the \emph{negation}, which maps an integer to its inverse, indicated by $-\iota$, or only by its sign $-$, in expressions. 

\begin{definition}
	The \textbf{inverse} (for concatenation) on $\Sigma_n^*$, denoted by $S^{-1}$ for $S\in \Sigma_n^*$, we define as $(S,T)^{-1}=(T^{-1},S^{-1})$ for $S,T\in \Sigma_n^*$, and $\lr{x}^{-1}=\lr{-x}$ for $x\in \Sigma_n$.
\end{definition}
We use the \emph{inverse} for concatenation to transform $\Sigma_n^*$ from a monoid into a \emph{free group}.
We observe $-\Rev(S)=S^{-1}$ for $S\in \Sigma_n^*$, where $\Rev$ is the reverse from \cref{def:reverse}. 
Hence, we use $-\Rev$ as the inverse on $\Sigma_n^*$. The inverse is an anti-morphism because it is a combination of two mappings, of which the reverse is an anti-homomorphism, and the negation is a homomorphism.

\begin{definition}
	A \textbf{substitution} is a morphism $T : \Sigma_n^* \to \Sigma_n^*$ that is \emph{expansive}, i.e., for at least one $x\in \Sigma_n$, we have $\|T\lr{x}\| \ge 2$.~\footnote{~A morphism is called expan\emph{ding} \cite[p.~9]{allshal} if for \emph{all} $x\in \Sigma_n$, we have $\|T\lr{x}\| \ge 2$.}
	Also, for every morphism $\s$ and substitution $T$, we have $T\s=\s T$.
\end{definition}

\subsection{Signed permutations}\label{sec:perms}
We recall that a \emph{signed permutation}\footnote{~Following Knuth \cite{Knuth}, we also use \emph{perm} to denote a signed permutation.} is a signed binary matrix, i.e., with elements $0,1,\text{ and }-1$, where each row and column has only one element distinct from $0$.
Clearly, such a permutation has determinant $\pm 1$ and therefore forms a member of the special orthogonal group.
\begin{lemma}\label{lem:sgndprm}
	If $\Omega=\omega(i,j)$ is the matrix of order $n$ of a signed permutation, then there is a permutation $\s$ of $\Sigma_n$, such that $\s(a)=-\s(-a)$ for $a\in \Sigma_n$.
\end{lemma}
\begin{proof}
Determine $\s(k);k=1,2,\ldots,n$, by the unique $1\le i \le n$ such that $\omega(i,k)\not=0$ and $\omega(j,k)=0$ for $j\not= i$, and define $\s(k)=i\ast\omega(i,k)$ for $1\le k\le n$.
Thus, we have $\sgn\big(\s(k)\big)=\omega(|\s(k)|,k)$, where $\omega(m,k)=0$ for $m\not=|\s(k)|$ and $1\le k\le n$.
Further, we define $\s(-k)=-\s(k)$ for $1\le k\le n$ and the condition is fulfilled. 
Then we get $\s(k)=\sgn(k)\ast\omega(|\s(k)|,|k|)$ for $k\in\Sigma_n$.
As an illustration for the above, see \cref{eq:sngndmtrx}.
\end{proof}
In Cauchy's two-line notation\footnote{~\cite[p.~94]{Wussing}, ``Cauchy used his permutation notation -- in which the arrangements are written one below the other, and both are enclosed in parentheses -- for the first time in 1815.''}, a permutation looks like 
$\begin{bmatrix} x & y & z & \cdots \\ a & b & c & \cdots \\\end{bmatrix}$, where the first row contains elements from the domain, and the second row contains their respective images.
A signed permutation is also a bijective morphism $\s$ on the alphabet $\Sigma_n$ with the property $\sigma(-x)=-\sigma(x)$ for $x\in \Sigma_n$.
Thus, we use the \emph{one-line notation} $\left[ \sigma(1), \sigma(2), \sigma(3),\ldots ,\sigma(n) \right]$, by which $\sigma$ is completely determined.\footnote{~Bj\"{o}rner et al. \cite[p.~246]{Bjorner2005} called this a \emph{window notation}. Section (8.1) \cite{Bjorner2005} is devoted to the properties of signed permutations and their group.}

Examples are the identity
$\iota=[1,2,3,\ldots,n]$ and its negation $-\iota=[-1,-2,\ldots,-n]$. 
As $[-2,4,-1,3]$ indicates the images of the four unit-vectors, the matrix becomes,
\begin{equation}\label{eq:sngndmtrx}
	[-2,4,-1,3]=
\begin{pmatrix}
	0&0&-1&0\\
	-1&0&0&0\\
	0&0&0&1\\
	0&1&0&0
\end{pmatrix}.
\end{equation}
\begin{lemma}
	Signed permutations
	$\s=[\s(1),\s(2),\ldots,\s(n)]$ and $\tau=[\tau(1),\tau(2),\ldots,\tau(n)]$, have as a product 
	\[\s\tau =\big[\s\big(|\tau(1)|\big)*\sgn\big(\tau(1)\big),\; \s\big(|\tau(2)|\big)*\sgn\big(\tau(2)\big),\;\ldots,\; \s\big(|\tau(n)|\big)*\sgn\big(\tau(n)\big)\big].\] 
\end{lemma}
 \begin{proof}
 	This product follows immediately from $\s(x)=\sgn(x)\ast\s(|x|)$ for $x\in\Sigma_n$.
 \end{proof}
For instance, $[-2,4,-1,3] [3,-1,4,-2]=[-1,-(-2),3,-(4)]=[-1,2,3,-4]$.
If we multiply from the right with e.g., $[1,3,2,4]$, we swap two values in \emph{positions} $2$ and $3$, as in\\ $[-2,4,-1,3][1,3,2,4]=[-2,-1,4,3]$.
Multiplying from the left swaps the corresponding \emph{values}, as we see in $[1,3,2,4][-2,4,-1,3]=[-3,4,-1,2]$.

From the matrix representation of signed permutations, we know the determinant equals $\pm 1$. The perms with matrices with positive determinants preserve the orientation of the vector space, and we call them \textbf{rotations}; those with negative determinants alter the orientation in the opposite one, and we call them \textbf{reflections}.
We now investigate whether a signed permutation is a reflection by observing its one-line notation.

\begin{definition}
Let $\s=[\s(1), \s(2),\ldots,\s(n)]$ be a signed permutation. Then, we define the \emph{number of negative items} \cite{Fire} by 
{\boldmath$\ng1$}$(\s)=\big|\{1\le i\le n: \s(i) < 0\}\big|$, and the \emph{number of inversions}\cite{Fire} by {\boldmath$\inv$}$(\s)=\big|\{(i, j): 1 \le i<j \le n,\; |\s(i)| > |\s(j)|\}\big|$.
\end{definition}
\begin{theorem}\label{th:detprms}
	$\det(\s)=-1^{\ng1(\s)+\inv(\s)}$.
\end{theorem}
\begin{proof}
	We know that the number of inversions, $\inv(\s)$, equals the number modulo $2$ of transpositions of two values in the one-line notation.
	Such a transposition, say $\s(i)$ and $\s(j)$, corresponds to swapping the two columns $i\text{ and }j$ in the corresponding matrix, which leads to an additional factor of $-1$ in the determinant.
	Multiplying a value in the one-line notation by $-1$ equals multiplying the corresponding column in the matrix with $-1$.
	Therefore, if we add the inversions and the minus signs in the one-line notation, we get the exponent of $-1$ in the determinant.
	We refer to the signed permutation $\mu=[2,3,4,\ldots,n,-1]$ as \textbf{the minimal rotation} because $\ng1(\mu)=1=\inv(\mu)$. 
\end{proof}

A sequence $S=\lr{s_1,s_2,\ldots}$ is generally not normalized (c.f.~\cref{def:normalized}), but we can easily construct a signed permutation to transform such an array into an isomorphic, normalized sequence.
For this, we need the first occurrence $|s_i|=k$ in the series for each $0<k\in\Sigma_n$, along with the $\sgn(s_i)$ of that first occurrence. In that order, these first occurrences for $0<k\in \Sigma_n$ constitute the inverse of the normalizing permutation of the original sequence.

\begin{definition}\label{def:char_perm}
	Given a sequence $S=\lr{s_1,s_2,\ldots}$ over $\Sigma_n$, define $i_k$ for $1\le k\le n$ such that $|s_{i_k}|=k$, and $|s_j|\not=k$ for $1\le j < i_k$, then we get 
$\big\{|s_{i_1}|,|s_{i_2}|, \ldots,|s_{i_n}|\big\}$.	
	This determines the \textbf{normalizing permutation} of $S$ to be $\s=\begin{bmatrix} s_{i_1} & s_{i_2} &\cdots& s_{i_n} \\1 & 2 &\cdots& n \\\end{bmatrix}$, or $\s^{-1}=[s_{i_1},s_{i_2},\ldots,s_{i_n}]$.
\end{definition}

A \emph{constant morphism} $\gamma:\Sigma_n\to \Sigma_n$ is such that for $x\in\Sigma_n$, we have $\gamma(x)=c$.
A \emph{constant substitution} $\gamma^*:\Sigma_n^*\to \Sigma^1_n$ is a mapping such that for $S\in\Sigma_n^*$, we have $\gamma(S)=\lr{c}$.
We use $\Sigma$ instead of $\Sigma_n$ if $n$ is evident from the context.

\subsection{Grid, direction, and isometry}\label{sub:gdi}

\begin{definition} 
	Let $\big\{u(1),u(2),\ldots,u(n)\big\}\subset \R^d$ be a set of vectors~\footnote{~We identify points, vertices, and vectors in $\R^d$.}, not necessarily independent, where $0<d\le n$, such that the vectors span $\R^d$. 
	Furthermore, we have $u(j)\not= \alpha\ast u(i)$ for $\alpha \in \R$ and $1\le i\not= j\le n$, i.e., every \emph{pair} of vectors is independent.
	The set $\Gamma_n=\left\{\sum\limits_{i=1}^n k_i\ast u(i) \right\}$ with $k_i\in \R$ and $|\{k_i\notin \Z\}|\le 1$, is called a \textbf{grid}.
	A grid has $2n$ \textbf{directions}, i.e., its generators and their negations
	$\{\pm u(k)\}$.
	Each direction with its opposite forms an \textbf{axis}.
	The generators of a grid $\Gamma_n$ relate to an alphabet $\Sigma_n$ by the mapping $u(k)\mapsto \lr{k}$ and $-u(k)\mapsto \lr{-k}$.
\end{definition}

Using the relation between the alphabet and the generators of the grid, we have an association between \emph{number sequences and fractal images}, i.e., subsets of the grid, as the title of this study suggests.

\begin{example}
	The most crucial grid we encounter is the cubic grid $\Z^d$ for $1\le d$, with the $2d$ directions $\lr{\pm 1},\lr{\pm 2},\ldots,\lr{\pm d}$, see \cref{fig:triangulargrid1} the left picture for the square grid in the plane.

\begin{figure}[H]
	\centering
	\includegraphics[scale=1]{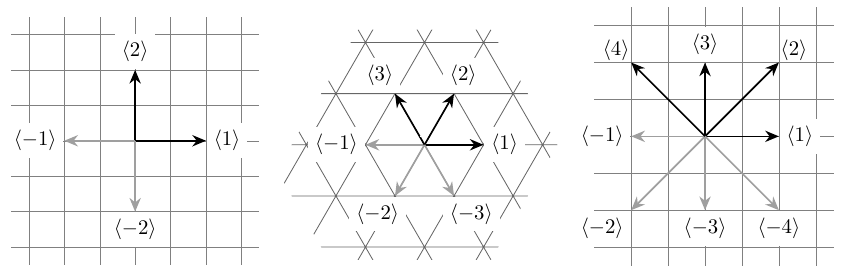}
	\caption{\small Square, triangular and square-diagonal grids, with directions indicated by integers.}
	\label{fig:triangulargrid1}
\end{figure}

The triangular grid 
is also essential.
Notice that a grid can have more generators than its dimension, as observed in the triangular and square-diagonal grids.
We determine a grid using a matrix in which the columns represent the generators in the appropriate order.
The grids in \cref{fig:triangulargrid1} are given by
$\begin{pmatrix}
	1 & 0 \\ 0 & 1 \\
\end{pmatrix}$, 
$\begin{pmatrix}
	1 & \frac{1}{2} & -\frac{1}{2} \\ 0& \frac{1}{2}\sqrt{3} & \frac{1}{2}\sqrt{3}\\
\end{pmatrix}$
 and 
$\begin{pmatrix}
	1 & 1 & 0 & -1\\ 0 & 1 & 1 & 1 \\
\end{pmatrix}$.

Contrary to a \emph{point lattice}, which consists of only vertices, a grid is a \emph{graph} with vertices and edges.
To the best of our knowledge, the only line fractals in higher dimensions are those on the cubic grid based on the cubic lattice.
\end{example}

\begin{example}
The following grids are associated with the triangular and square-diagonal grids: the \emph{tri-hexagonal grid}, \emph{hexagonal} (honeycomb) grid, and \emph{truncated square} grid, see \cref{fig:hc_trsq_grd}.
\begin{figure}[H]
	\centering
	\includegraphics[scale=.7]{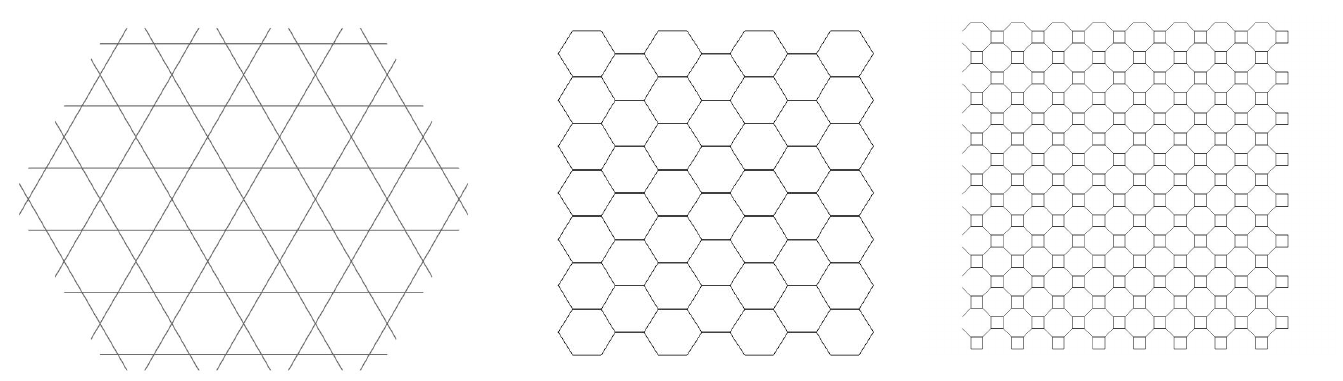}
	\caption{\small Tri-hexagonal, Honeycomb, and Truncated square grids.}
	\label{fig:hc_trsq_grd}
\end{figure}
We use directions from the triangular and square-diagonal grids, but with restrictions, as displayed in \cref{fig:hc_trsq_grd_2}.

\begin{figure}[H]
	\centering
	\includegraphics[scale=0.60]{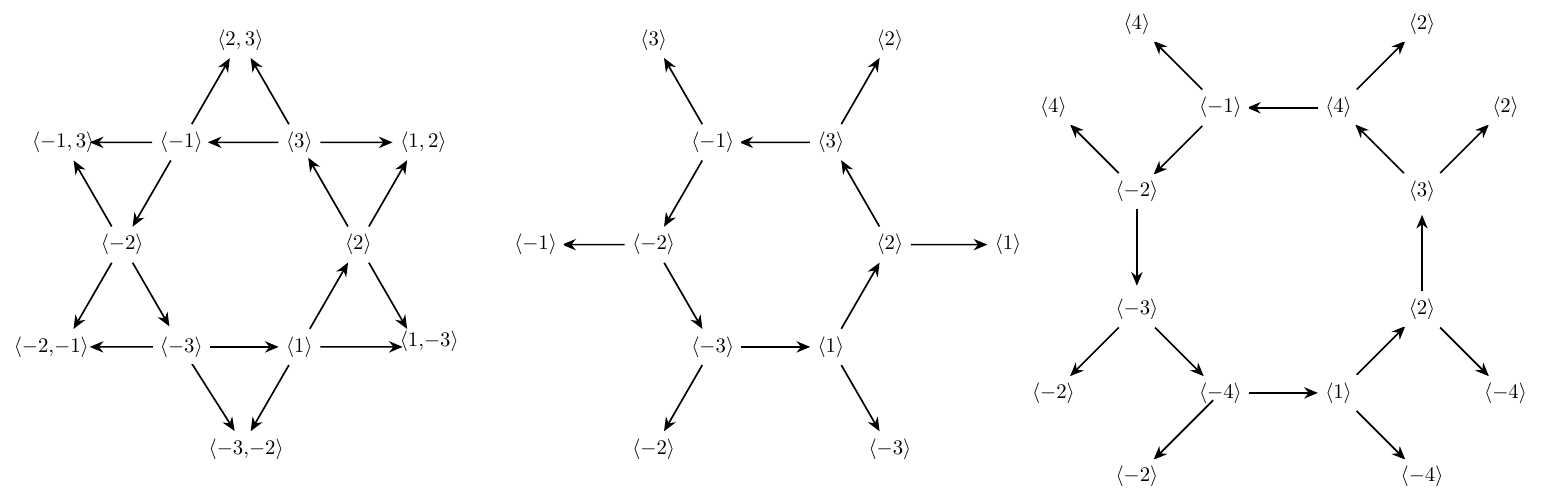}
	\caption{\small Successor directions in the tri-hexagonal, honeycomb, and truncated square grids. The vertices indicate the incoming directions.}
	\label{fig:hc_trsq_grd_2}
\end{figure}

In \cref{fig:hc_trsq_grd_2}, the edges that can follow a specific direction are shown in each of the central polygons. In the tri-hexagonal grid, three directions follow one direction. By contrast, there are two possible directions at a vertex of the honeycomb or the truncated square grid.
\end{example}

Generally, a \emph{fractal} is represented as a geometrical figure, where the approximants undergo shrinking, such as 
$F_{n+1}=\varphi\ast T(F_n)$ with $0<\varphi<1$ and an (expansive) substitution $T$.
Therefore, we define the fractal as $F=\lim\limits_{n\to\infty}F_n$.
Our infinite sequence, with the integers interpreted as directions in a grid, becomes a geometrical object of infinite size, where the size or length of each approximant is less than that of the next.

In our approach, the approximants of a fractal are \emph{finite curves}, i.e., directed graphs with all vertices of degree two, except for the first and the last one, which have both degree one.
Therefore, an \emph{entry (point)} is the first vertex on the curve, and we depict it by $\circ$, an \emph{exit (point)} is the last vertex on the curve, and we depict this by $\bullet$.

A fractal, being an infinite limit of finite curves, has a single entry.
Therefore, we refer to a fractal as a \emph{curve} and replace the term ``approximant'' with \emph{$k$-curve} or \emph{curve of level $k$}.
\label{def:orientation}
The \emph{orientation} of a $k$-curve is the vector directly from entry to exit.

We define a vertex as the $0$-curve, or a curve without edges \cite{allshal}, denoted by $\epsilon=\lr{}$.
Most of the curves we generate from a substitution that uses the $1$-curve, i.e., the first approximant, as the start.

\begin{remark}
	A sequence represents subsequent edges; therefore, an edge and its opposite are more than a single vertex, and they graph-wise indicate two successive edges in opposite directions.
	Hence, we do \emph{not} use annihilation, i.e., we do \emph{not} identify $\lr{}$ with $\big(S,-\Rev(S)\big)$ or $\lr{a,-a}$.
\end{remark}

\begin{definition}
	An \textbf{isometry} is a distance-preserving transformation on the grid. If necessary, we denote two isometric sets by $A\cong B$.
\end{definition}

Notice that an isometry also \emph{preserves the angles} between vectors.
Whether a signed permutation is an isometry depends on the difference in the length of the generators, for instance, the square-diagonal grid in \cref{fig:triangulargrid1}.

The inverse $-\Rev$ of an (only finite) curve is a mapping from the curve to itself, with entry and exit swapped, the ordered multiset of edges reversed, and the direction of each edge reversed. 
The reverse $\Rev$ only reverses the multiset of edges; hence, the entry and exit remain fixed.
Therefore, if $S=\lr{e_1,e_2,\ldots,e_n}$ are the edges of a curve, then $-\Rev(S)=\lr{-e_n,\ldots,-e_2,-e_1}$ and $\Rev(S)=\lr{e_n,\ldots,e_2,e_1}$.
See Intermezzo 1 on page \pageref{int:intermezzo1}.

\subsection{Fractals and approximants}\label{sub:frctlapprox}

Fractals are difficult to define. Mandelbrot, who coined the term \cite{Mbrt}, described them as
``\emph{(...)~a rough or fragmented geometric shape that can be split into parts, each of which is - at least approximately - a reduced-size copy of the whole.}''
Falconer \cite{Falc} stated ``\emph{My personal feeling is that the definition of a `fractal' should be regarded in the same way as a biologist regards the definition of `life'.}'' He refers to a fractal as an object with five different, explicit properties, not all of which he describes with sufficient precision.
Finally, Kimberling \cite{Kimb} wrote ``\emph{A search of \cite{OEIS} for `fractal sequence' reveals that in recent years, different kinds of
	sequences have been called `fractal' and what many of them have in common is that they are SCSs.}'' (= Self-Contained Sequences).

We define a fractal sequence as the limit of (an infinite number of) mutually related, finite sequences of increasing length, resulting in an infinite sequence.

\begin{definition}
	A \textbf{fractal} is the limit of an ordered, infinite set of extending sequences, i.e., each sequence of finite length is larger than the length of the previous sequence, which we call \textbf{approximants}. Each one is a concatenation of the images of the former approximants, usually only the previous one.
	Therefore, if we have a countable infinite set of finite sequences $\{S_1,S_2,\ldots\}$ and a set of morphisms $\big\{\alpha_{i,k}:\Sigma^*\to \Sigma^*; i=1,2,\ldots,n; n\in \N \big\}$, and substitutions $T_k:\Sigma^*\to \Sigma^*$, such that $S_{k+1}=T_k\big(S_1,\ldots,S_k\big)=\big(\alpha_{1,k}(S_{1,k}),\alpha_{2,k}(S_{2,k}),\ldots,\alpha_{n,k}(S_{n,k})\big)$, for $S_{j,k}\in\{S_1,\ldots,S_k\};j=1,2,\ldots,n$,
	then the fractal is $S=\lim\limits_{k\to\infty}S_k \in \Sigma^{\N}$, the set of (right-)\emph{infinite} sequences.
	
	A fractal is called \label{def:slfsmlr} \textbf{self-similar} if $S_{j,k}=S_k$ and $\alpha_{j,k}=\alpha_j$ for $k>0$ and $j=1,2,\ldots,n$, thus 
	$S_{k+1}=T\big(S_k\big)=\big(\alpha_1(S_k),\alpha_2(S_k),\ldots,\alpha_n(S_k)$.
	Henceforth, we assume all our curves to be self-similar.	
	Furthermore, without loss of generality, we assume the fractals to be \textbf{extending}\label{df:extending}, i.e., with identity $\alpha_1=\iota$ such that $S=\big(S_k,\ldots\big)$ and $S_k$ is a prefix of $S_{k+1}$ for $k>0$.
	We generally identify a fractal with one of its approximants.	
\end{definition}
For a self-similar fractal, we write
\begin{equation}\label{eq:substbymorph}
	T\big(S_k\big)=\big(S_k, \alpha_2(S_k), \ldots, \alpha_n(S_k)\big) 
	\text{ and denote this by }
	{T=[\iota, \alpha_2,\ldots,\alpha_n]}~\footnotemark
\end{equation}
 \footnotetext{~For $\s\in \Omega$, the group of morphisms on $\Sigma^*$, a substitution $T:\Sigma^*\to\Sigma^*$ implies a dual substitution ${}^m T:\Omega\to\Omega$, by ${}^m T(\s)=T\; \s$.}

\subsection{Ventrella's flags}\label{sub:ventrflgs}
\begin{figure}[H]
	\centering
	\includegraphics[scale=1.2]{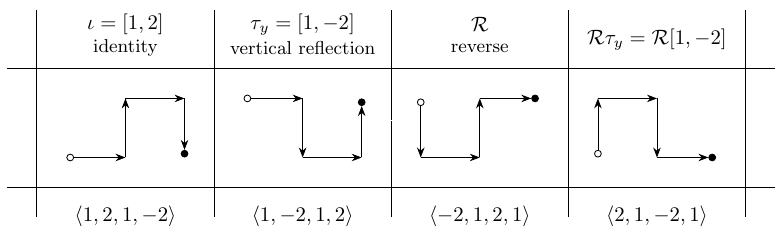}
	\caption{\small Different directions of a $1$-curve.}
	\label{fig:ventrllflags}
\end{figure}

Suppose we have a fractal whose $1$-curve is identical to the first image in \cref{fig:ventrllflags}, under ``identity.''
If this is the image under the substitution of a horizontal unit line segment, then we investigate what the images of the other line segments are, either horizontal or vertical.  
We can choose the transformations of the first curve, with similar entry and exit, as in the rest of \cref{fig:ventrllflags}.

\begin{figure}[H]
	\centering
	\includegraphics[scale=1.15]{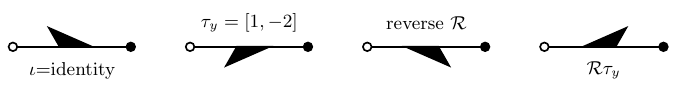}
	\caption{\small Ventrella's flags and its isometries.}
	\label{fig:ventrllflags2}
\end{figure}

Ventrella \cite{Ventrella}, who was not a mathematician but an artist, suggested a one-sided, flag-like arrow \cite{VentrellaTree} to indicate different oriented edges (see the center drawing of the first row of \cref{fig:ventrll1crvs}), which led to various images of those edges.
We place the flag at the center and create the drawings in \cref{fig:ventrllflags2}, which correspond with those in \cref{fig:ventrllflags}.

\begin{figure}[H]
	\centering
	\includegraphics[scale=1.2]{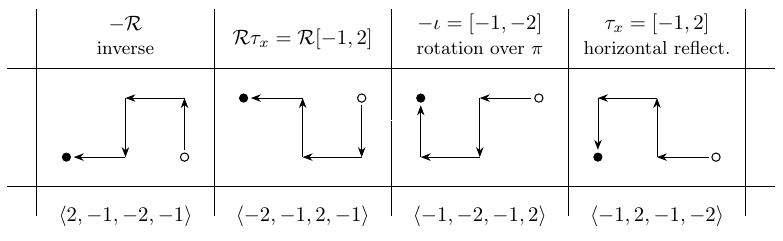}
	\includegraphics[scale=1.15]{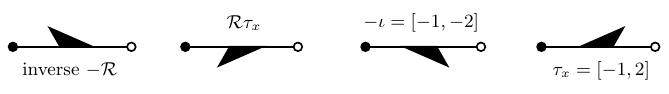}	
	\caption{\small Similar drawings as \cref{fig:ventrllflags} and \cref{fig:ventrllflags2}, other isometries swapping entry/exit and directions.}
	\label{fig:otherflags}
\end{figure}

In \cref{fig:otherflags}, we observe similar figures as in \cref{fig:ventrllflags} and \cref{fig:ventrllflags2}, with their corresponding transformations, but we swap the directions, just like the entry and exit.
This figure completes the list of all the isometries of the original figure, except the (infinite number of) rotations.

\setlength{\fboxsep}{3mm}
\noindent\rule{2mm}{0mm}\framebox[16cm]{
	\parbox{15.5cm}
	{\textbf{Intermezzo 1}\label{int:intermezzo1}\\	
					
		\begin{minipage}[b]{7.5cm}
			There is an issue in terms of the difference between reverse, negation, and rotation over $\pi$, since we can swap the order of the edges, swap the direction of each edge, in which case the entry and exit are swapped as well, or both.
			
			In figure \cref{fig:ventrllflags3}, we illustrate different ways to reverse a directed graph with entry and exit.
			
			As we can see, $-\iota=\text{\sl negate}$ produces a rotation over $\pi$ without changing the entry and exit, although they changed their relative position. 
			$-\Rev=${\sl inverse}, which annihilates the original by reversing everything, the order of edges, the directions of edges, and consequently, the entry and exit.
		\end{minipage}
		\begin{minipage}[b]{7.5cm}
			\begin{figure}[H]
				\centering
				\includegraphics[scale=1.1]{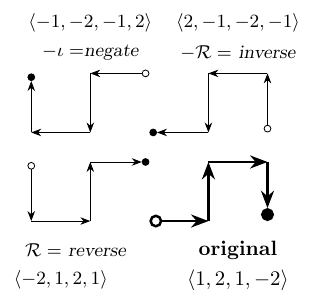}
				\caption{\small Rotate by reverse or negate.} \label{fig:ventrllflags3}
			\end{figure}
		\end{minipage}
		However, $\Rev=\text{\sl reverse}$ keeps the entry and exit fixed, by swapping the directions of the edges, so it appears to be the rotation over $\pi$ of the \emph{inverse}. 
				
		When we use a rotated $k$-curve in the build-up of the next version, we can reverse the order of the edges without swapping the entry and exit, but we can also switch the edges themselves.

		We refrain from using the phrase ``rotate over $\pi$,'' and use one of the three different formulas, $-\Rev$, $-\iota$, or $\Rev$, for the sake of clarity.}}

\vspace{0.5\baselineskip}
We can decorate the original $1$-curve by choosing one of the flags from \cref{fig:ventrllflags2} or \cref{fig:otherflags} for each edge to determine in which image of the $1$-curve this edge can be substituted. 
The (open) question remains: how many different, i.e., non-isometric curves can we construct using only normalized curves?

\subsection{Representing fractals concisely}\label{sub:reprfrun}
One of our objectives is establishing an encyclopedia of normalized fractals as an independent database. However, such an encyclopedia could be considered as a subset of \href{https://oeis.org/}{OEIS}.
\begin{definition}\label{df:seqsort}
	The \textbf{ordering} of the sequences in the database is such that for the normalized sequences $S=\Lr{s(1),s(2),\ldots,s(k),{s(k+1)},\ldots}$ and $S'=\Lr{s(1),s(2),\ldots,s(k),\mathbf{s'(k+1)},\ldots}$ with $s(k+1)\not=s'(k+1)$, we have 
	\[S\prec S' \Longleftrightarrow |s(k+1)<|s'(k+1)|\text{ or }\;0<s(k+1)=-s'(k+1)\]
	Notice that the absolute values we sort non-descending, and those numbers with equal absolute values, we sort positives first, like $1,-1,2,-2,3,-3,\ldots$.
\end{definition}
The \textbf{bold} number in a sequence is the first number that makes this sequence differ from, and larger than the previous one in the encyclopedia.

We describe the fractal by the first twenty or so items of the sequence, its index in \href{https://oeis.org}{OEIS}, if any, and subsequently give the alphabet, start sequence, substitution, and grid with its generators, and finally the figure of the geometric fractal.
See the example below of the first sequence (\cref{sub:Dkkng Flwsnk}) in the database.

\begin{description}
\setlength{\itemsep}{0ex}
	\item[B.3 Dekking's Flowsnake] 
	\item[sequence:] $\lr{\mathbf{1},1,2,-1,2,1,2,-1,-1,2,1,1,1,2,1,-2,-2,-1,-2,-2,1,2,1,-2,-2,1,\ldots}$
	\item[in \href{https://oeis.org}{OEIS}:] \seqnum{A356112}.
	\item[alphabet:] $\Sigma=\{1,2\}$
	\item[start sequence:] $\lr{1}$
	\item[substitution:]
	Let $\tau=[1,-2]$ and $\mu=[2,-1]$, then, conform \cref{eq:substbymorph} on page \pageref{eq:substbymorph} we get\\
	$\rule{-5mm}{0mm}T(\iota)=(\iota,\iota,\mu\tau,-\tau,\mu,\iota,\mu\tau,-\tau,-\iota,\mu\tau,\iota,\iota,\tau,\mu,\tau,-\mu,-\mu,-\tau,-\mu,-\mu\tau,\tau,\mu,\iota,-\mu\tau,-\mu\tau)$. 
	\item[grid:] The square, plane grid.
	\item[generators:] $(1,0) ; (1,0)$
	\item[figure:] The $1$-curve (first row left), two $2$-curves (anti-diagonal), the $3$-curve (below right).
\end{description}

\begin{figure}[H]
	\begin{center}
		\includegraphics[scale=0.31]{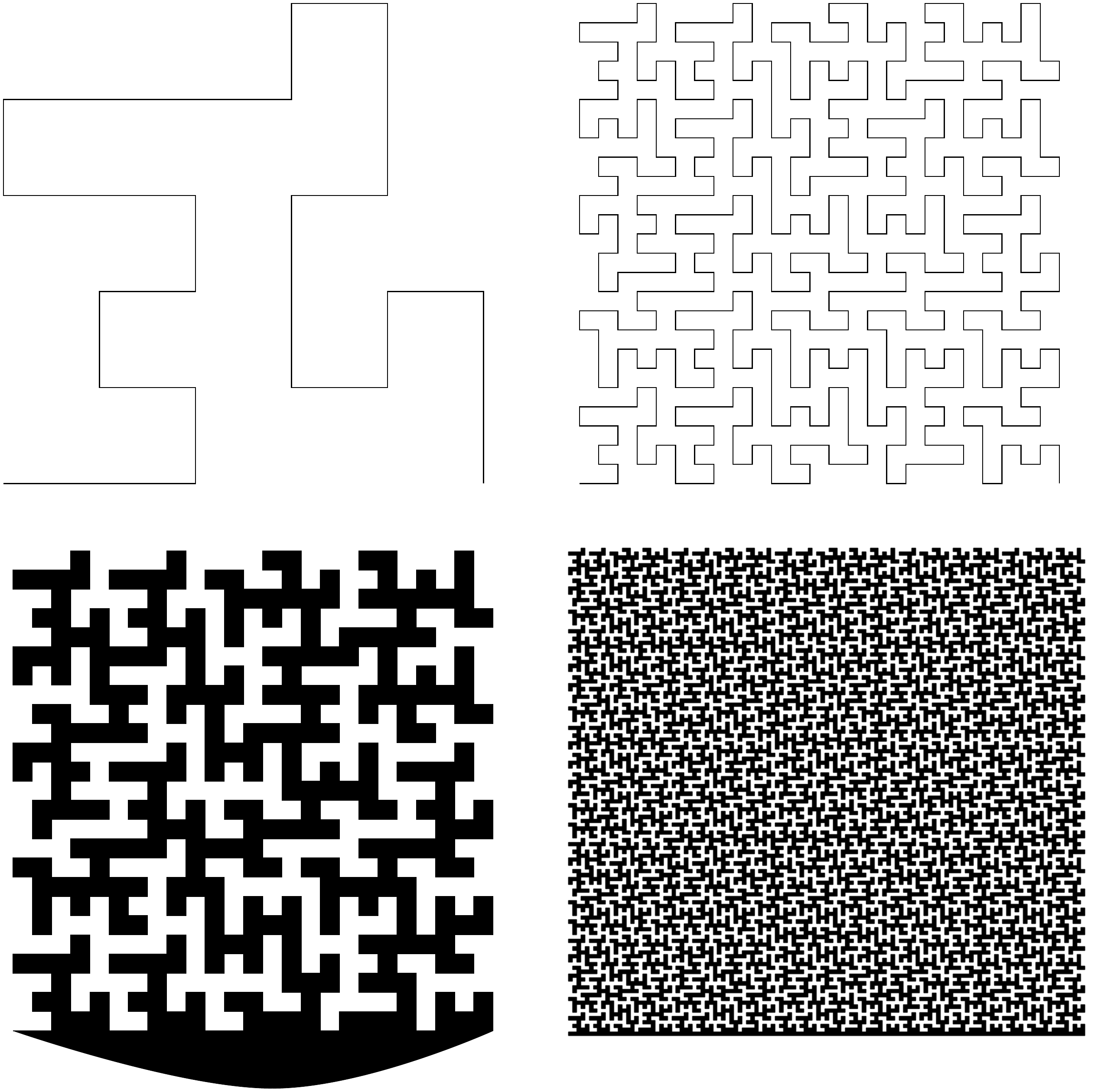}
		\caption{\small{First row is drawings of the $1^{\text{st}}$ and $2^{\text{nd}}$ approximants. The second row is the $2^{\text{nd}}$ and $3^{\text{rd}}$ approximants, separating the space into two parts, black and white, both with tree-like structures.}}
	\end{center}	
\end{figure}

\section{Examples}\label{sec:exmpls}
The following examples highlight different aspects of representing fractals as integer sequences. 

\subsection{Hilbert's original curve}\label{sub:hlbrt2d}
\begin{figure}[H]
	\centering
	\includegraphics[scale=1.1]{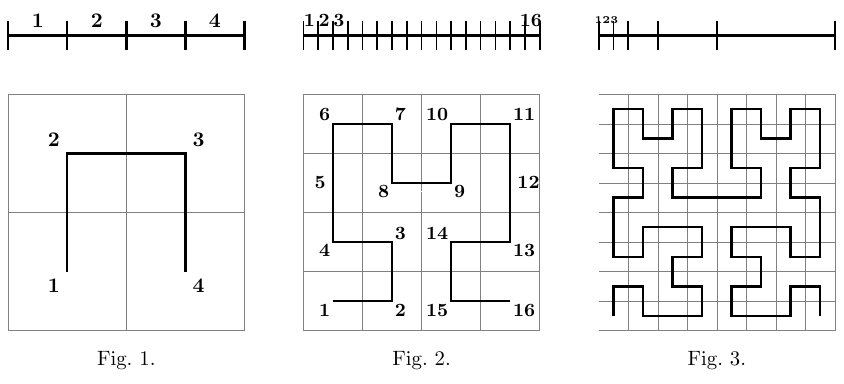}
	\caption{\small A copy of Hilbert's original drawings of his first, second, and third approximants.} 
	\label{fig:hilbert}
\end{figure}

 In this subsection, we study one of the oldest and most famous fractals, the curve Hilbert presented in his two-page paper \cite{Hilbert}, with the drawing he depicted and we redrew in \cref{fig:hilbert}, as the primary explanation.

\begin{figure}[H]
	\centering
	\includegraphics[scale=0.60]{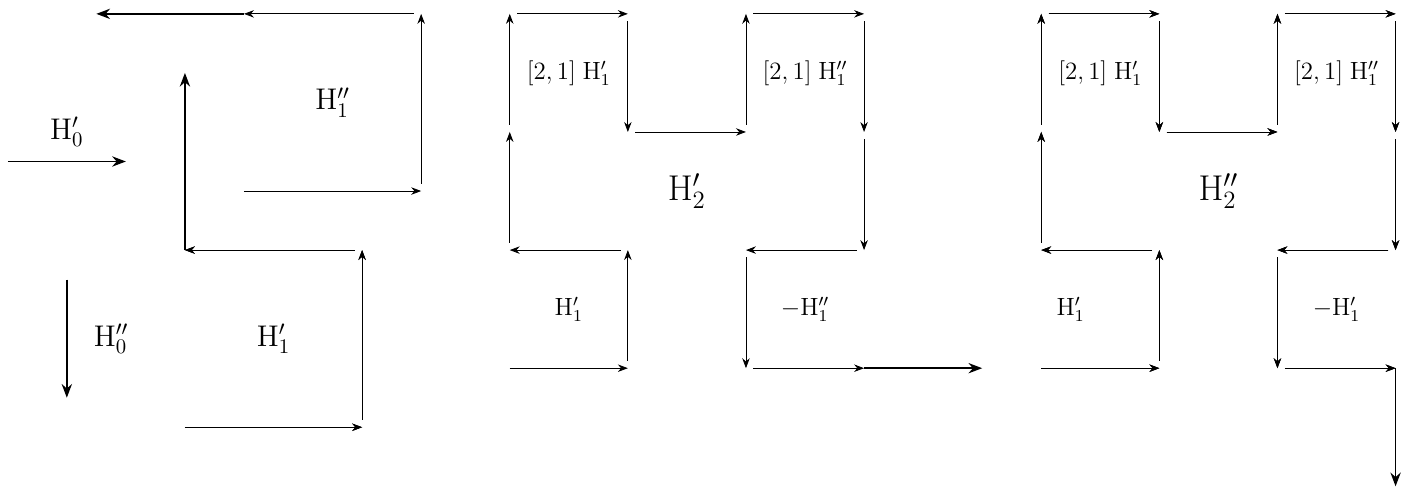}
	\caption{\small Hilbert's normalized and extending, $0$, $1$, and $2$-curves $H'_k$ and $H''_k$, with the extra \emph{exit edge} in the direction of, or orthogonal to the orientation, respectively.}
	\label{fig:hilbert4b}
\end{figure}

We propose a new and thorough way of generating the fractal sequence discussed in the introduction.
The Hilbert approximants are part of the square grid, as shown in the left drawing of \cref{fig:triangulargrid1}.
We make some preparations first, by enhancing the Hilbert $k$-approximant into two versions, both with an extra edge after the exit, one, H$'_k$, in the direction of the orientation\footnote{~Remember that the orientation of $k$-approximant, or $k$-curve, is the direction from entry to exit.}; the other one, $'$H$_k$, with the extra edge orthogonal to that orientation, see \cref{fig:hilbert4b}.

Let the signed permutation on the square grid $\tau_{d}=[2,1]$ be the \emph{diagonal} reflection in the line $y=x$.
 
\begin{observation}\label{ob:hlbrt1stsubst}
	The substitution $T$, which generates the normalized and extending sequence that represents the Hilbert curve, we define by
\begin{multline}\label{eq:hilbert1}
	H'_{k+1}=T\big(H'_k\big) =\Big(H'_k,\tau_{d}\big(H'_k\big), \tau_{d}\big(H''_k\big), -H''_k\Big)\;;\\
	H''_{k+1}= T\big(H''_k\big) =\left(H'_k,\tau_{d}\big(H'_k\big), \tau_{d}\big(H''_k\big),-H'_k\right).
\end{multline} \
	This substitution produces the sequence \\
	$\lim\limits_{k\to\infty} H'_k = \lr{1,2,-1,2,2,1,-2,1,2,1,-2,-2,-1,-2,1,1,2,1,-2,1,1,2,-1,2,1,2,-1,-1,\ldots}$,\\ known in \href{https://oeis.org}{OEIS} as \seqnum{A163540}, when we replace $1,2,-1,-2$ by $0,1,2,3$, respectively.
\end{observation}	
\begin{proof} 
	Since Hilbert himself gave only the three drawings of \cref{fig:hilbert} to characterize his curve, we can only show that our substitution generates the corresponding sequences.
	For $k=0$ we have $H'_0=\lr{1}$ and $H''_0=\lr{-2}$, thus we get for 
	$H'_1=(H'_0,[2,1]H'_0,[2,1]H''_0,-H''_0)=\lr{1,2,-1,2}$ and 
	$H''_1=\lr{1,2,-1,-1}$, and, apart from their exit edge, both coincide with Hilbert's Fig.~2.\\
	Then, $H'_2=(H'_1,[2,1]H'_1,[2,1]H''_1,-H''_1)=\lr{1,2,-1,2,2,1,-2,1,2,1,-2,-2,-1,-2,1,1}$ and\\ 
	$H''_2=(H'_1,[2,1]H'_1,[2,1]H''_1,-H'_1)=\lr{1,2,-1,2,2,1,-2,1,2,1,-2,-2,-1,-2,1,-2}$.
	Notice that only the exit edge between $H'_k$ and $H''_k$ is different.
\end{proof}
An advantage of these enlarged Hilbert curves is that they can be generalized to higher dimensions to produce high-dimensional (Hilbert) curves with unique properties \cite{BH}.

This Hilbert sequence we can also construct by a simple substitution on $4$ axes  represented by $\Sigma_4=\{\pm 1, \pm 2, \pm3, \pm 4\}$, and then ``projecting'' this sequence to the two axes represented by $\Sigma_2=\{\pm 1, \pm 2\}$.
\begin{equation*}
	T=\begin{cases}
		1 &\to \lr{ 1,2,3,4}\\
		2 &\to \lr{ 2,1,-4,-3} \\
		3 &\to \lr{ 2,1,-4,-2} \\
		4 &\to \lr{ -1,-2,-3,1}\\
	\end{cases}
	\quad\; T(-x)=-T(x)\text{ for }x\in \{\pm 1, \pm 2, \pm 3,\pm 4\}.
\end{equation*}
Note we found no isometry $\s$ on the alphabet $\Sigma_4$ such that $T\big(\s(x)\big)=\s\big(T(x)\big)$ for all $x\in \Sigma_4$.\\
The \emph{$4$-axes Hilbert sequence} is\\ $\lr{1,2,3,4,2,1,-4,-3,2,1,-4,-2,-1,-2,-3,1,2,1,-4,-3,1,2,3,4,1,2,3,-1,-2,\ldots}.$\\
After replacing $\mp 3$ by $\pm 1$ and $\pm 4$ by $\pm 2$, we get our Hilbert sequence\\
$\lr{1,2,-1,2,2,1,-2,1,2,1,-2,-2,-1,-2,1,1,2,1,-2,1,1,2,-1,2,1,2,-1,-1,-2,\ldots}.$

\begin{figure}[H]
	\centering
	\includegraphics[scale=0.50]{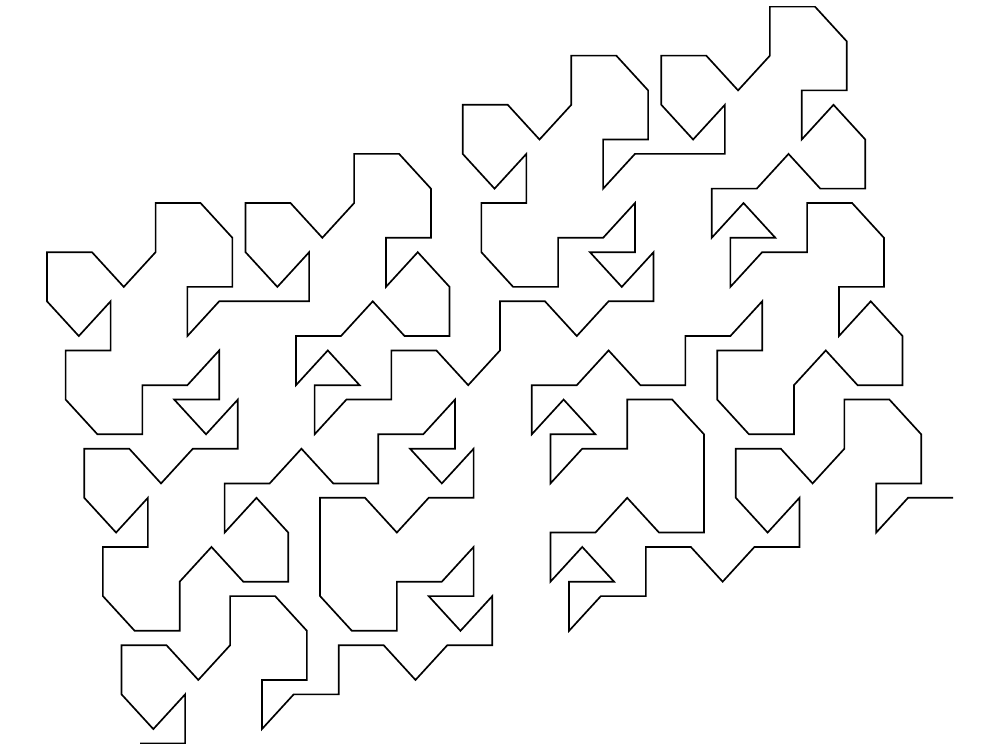}\quad\quad
	\includegraphics[trim=0 -0.8cm 0 0, scale=1]{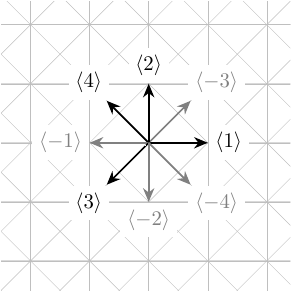}
	\caption{\small The $4$-axes Hilbert fractal on the renumbered $8^\text{th}$-root grid.} 
	\label{fig:hilbert4ext}
\end{figure}

\subsection{The \texorpdfstring{$\beta,\Omega$}{beta, Omega} curves}
\begin{figure}[H]
\begin{minipage}[t]{\textwidth}
	\centering
	\includegraphics[scale=0.7]{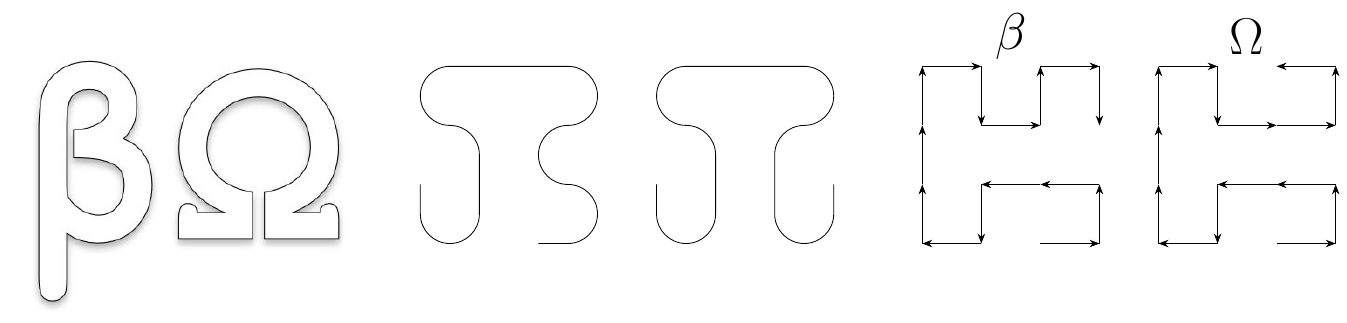}
	\caption{\small On the left are the letters to justify the naming, in the middle is the corresponding artist's impression, and on the right are the normalized $\beta$ and $\Omega$ $2$-curves.~\protect\footnotemark}
	\label{fig:btOm1} 
\end{minipage}
\end{figure}

In this section, we construct a pair of intertwining curves, which can be considered the nephews of the Hilbert curve.
We introduce the $\beta,\Omega$-curves, as referred by their inventor \cite{Wierum}, see \cref{fig:btOm1}.

\footnotetext{~Notice that there can be another non-isomorphic normalization of the $\beta$-curve if you reverse the $\beta$-curve first.}

\begin{figure}[H]
		\centering
		\includegraphics[scale=0.85]{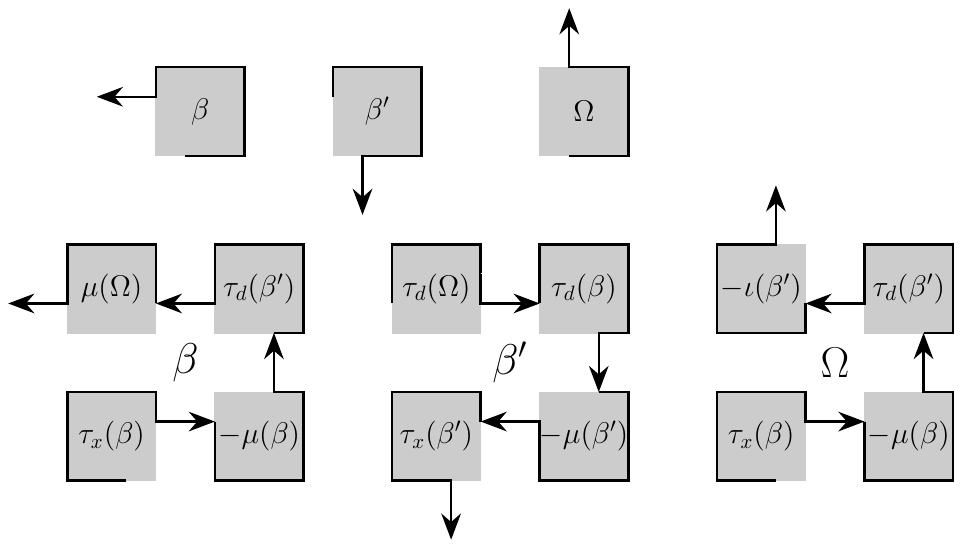}
		\caption{\small The first row shows the two general forms of the $k$-curves of type $\beta$, and the general form of the $k$-curves of type $\Omega$. We display the substitutions for the next generation underneath. See \cref{sc:Cayley} for the dihedral group D4 of transformations of the square grid, where the isometries $\pm\iota$, $\tau_x$, $\tau_d$, and $\mu$ are described.}
		\label{fig:btOm2} 
\end{figure}

The approximants of the Hilbert curve have entry and exit on the \emph{vertices} of the surrounding square, whereas the $k$-curves for $k\ge 2$ of the $\beta$- and $\Omega$-curve have entry and exit on (approximately) \emph{one-third of different edges} of that square.
Therefore, the general picture of a $k$ curve for the $\beta$ and $\Omega$-curves should look like the upper row of \cref{fig:btOm2}.
As we did before with the Hilbert approximants, we added an extra \emph{exit edge}.

\begin{figure}[H]
	\centering
	\includegraphics[scale=0.5]{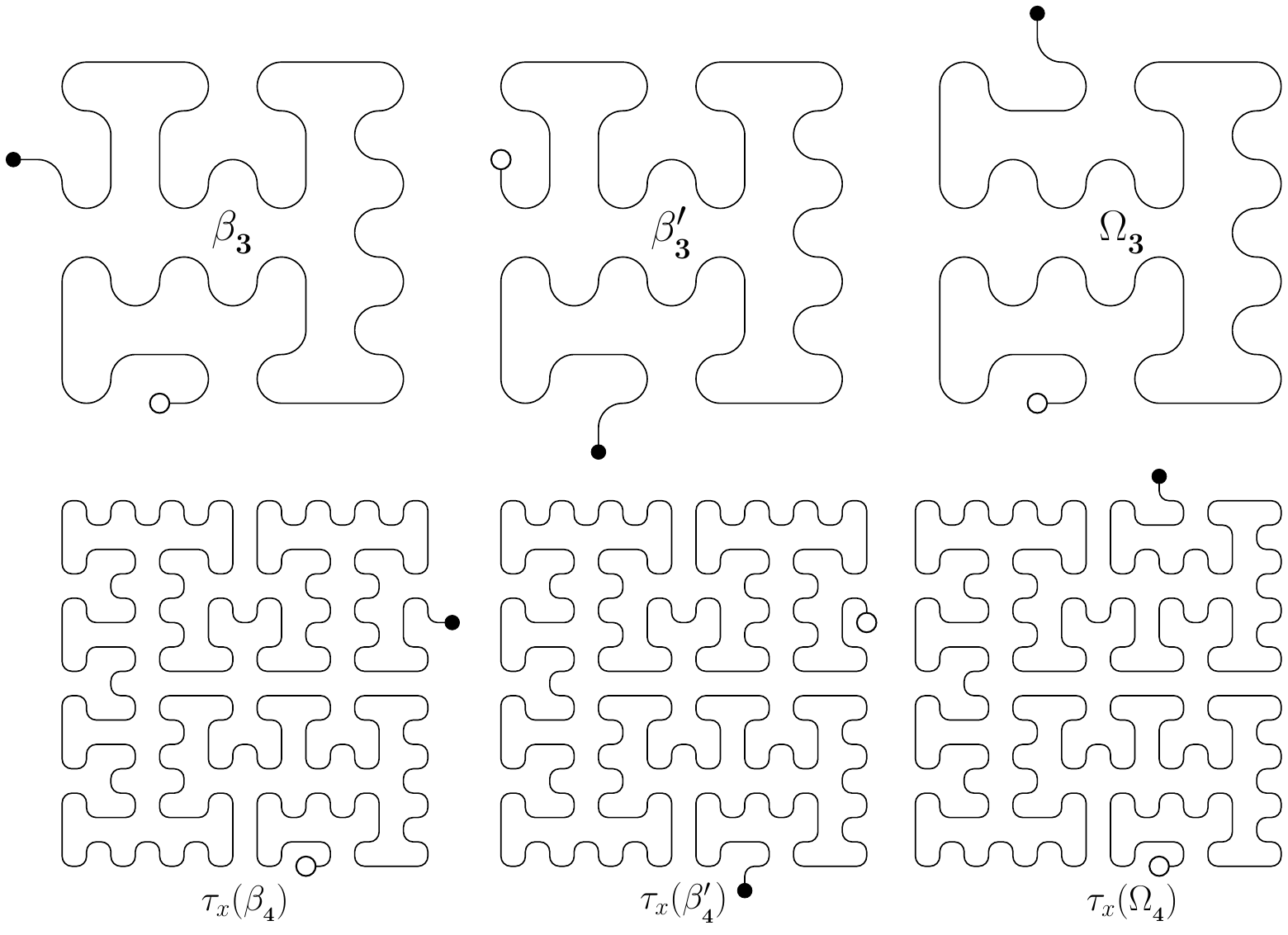}
	\caption{\small Upper and lower row give the $3$- and $4$-curves of the $\beta$ and $\Omega$ fractals, respectively, with entries and exits. 
		The $\Omega$-approximants follow the drawings in \cite{HarmHilb}.}
	\label{fig:betom23} 
\end{figure}

\noindent\fbox{\parbox{15.5cm}
	{\textbf{Intermezzo 2}\label{int:intermezzo2}
		
	 We require the extra curve $\beta'$ because of the asymmetry of the $\beta$-curve. $\beta'$ is approximately the inverse of the $\beta$-curve, as one can observe from the two first figures of the upper row of \cref{fig:betom23}.
	 Or, if $\beta=\Lr{s(1),s(2),\ldots,s(n-1),s(n)}$, then $\beta'=\Lr{-s(n-1),\ldots,-s(2),-s(1),-\tau_d\big(s(n)\big)}$.
	 
	 Likewise, we could redefine the $\Omega$-curve: if $\beta=(\alpha_1,\alpha_2,\alpha_3,\alpha_4)$ consists of four sub-curves of equal sizes, then $\Omega_k=\big(\alpha_1,\alpha_2,\alpha_3,(-\iota)^{k+1}\tau_d(\alpha_4)\big)$.	
	 We neglect these more complicated isometries and prefer the alternate curves $\beta'$ and $\Omega$.	
	}}\\

Therefore, we start with the $1$-curves $\beta_1=\lr{1,2,-1,-1}; \beta_1'=\lr{1,-2,-1,-2}\text{ and }\Omega_1=\lr{1,2,-1,2}$, conform to the first row of \cref{fig:btOm2}.\footnote{~In \cref{fig:btOm2}, a non-normalized version of $\beta'$ is given and used further because we do not use it at the start of a next approximant.}

As shown in the bottom row of \cref{fig:btOm2}, the next generation of the $\beta,\Omega$ curves is not normalized if the former one is. 
Thus, we apply the horizontal reflection $\tau_x$ alternately.
To normalize the $k$-curves, we combine the constructions in the lower part of \cref{fig:btOm2}, and obtain the following substitutions for $k\ge 1$:
\begin{alignat*}{3}
	T_{\beta}(k+1)&=\tau_x^k (\beta_{k+1} ); \quad \beta_{k+1}&=&\Big(\tau_x(\beta_k),-\mu(\beta_k),\tau_d(\beta'_k),\mu(\Omega_k)\Big) \\
	T_{\beta'}(k+1)&=\tau_x^k(\beta'_{k+1}); \quad \beta'_{k+1}&=&\Big(\tau_d(\Omega_k),\tau_d(\beta_k),-\mu(\beta'_k),\tau_x(\beta'_k)\Big)\\
	T_{\Omega}(k+1)&=\tau_x^k(\Omega_{k+1}); \quad
	\Omega_{k+1}&=&\Big(\tau_x(\beta_k), -\mu(\beta_k),\tau_d(\beta'_k), -\iota(\beta'_k)\Big)
\end{alignat*}
Refer to \cref{sc:Cayley} for the dihedral group D4 of transformations of the square grid. 
For $k=2$, we get the $2$-curves, as in the right-hand side part of \cref{fig:btOm1}:	
\begin{align*}
	\beta_2&=\lr{1,2,-1,-1,-2,-1,2,2,2,1,-2,1,2,1,-2,1},\\
	\beta'_2&=\lr{2,-1,-2,-1,2,-1,-2,-2,-2,1,2,1,1,-2,-1,-2},\\
	\Omega_2&=\lr{1,2,-1,-1,-2,-1,2,2,2,1,-2,1,1,2,-1,2}.\\
 \end{align*}
The $\beta$ and $\Omega$ curves are also mutually dependent, similar to the Hilbert curve with its extended approximants.
Their two approximants only differ in the last constituent.

\begin{figure}[H]
	\centering
	\includegraphics[scale=0.55]{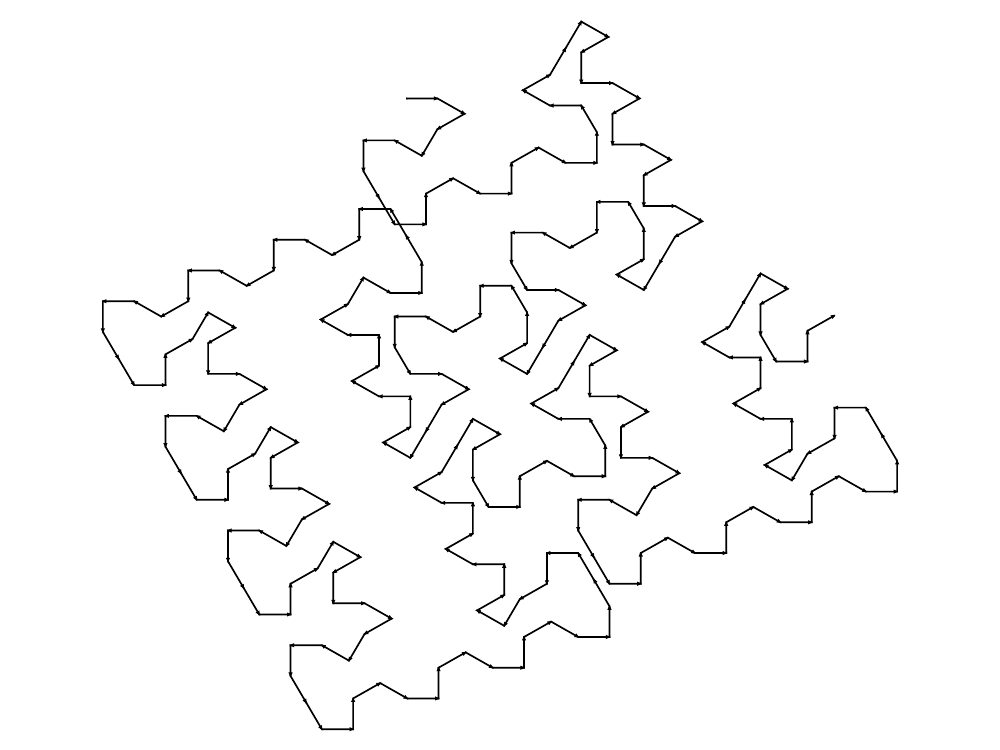}
	\includegraphics[trim=-0.3cm -0.8cm 0 0, scale=1.1]{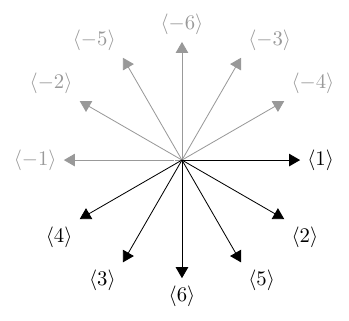}
	\caption{\small The $6$-axes $\beta\Omega$ fractal on the renumbered $12^\text{th}$-root grid.} 
	\label{fig:6dimcrv}
\end{figure}

As $\lim\limits_{k\to\infty} \beta_k=\lim\limits_{k\to\infty}\Omega_k$, the corresponding sequence for $\beta$ suffices and equals \\
$\lr{1,2,-1,-1,-2,-1,2,2,2,1,-2,1,2,1,-2,1,2,1,-2,-2,-1,-2,1,1,1,2,-1,2,1,2,\ldots}$.\\
There is a number-substitution possible, albeit not very simple. 
The one we find counts three variables in each direction, so it represents a curve on a $6$-axes grid, \cref{fig:6dimcrv}. The substitution is
\begin{equation*}
	T=\begin{cases}
		1 &\to\lr{ 1,2,3,4}\\
		2 &\to\lr{ -2,-1, 5,6}\\
		3 &\to\lr{ 6,1, -5,-3} \\
		4 &\to\lr{ 2,1,-5,-3}\\
		5 &\to\lr{ -4, 2,3,5} \\
		6 &\to\lr{ 1,2,3, 5} .
	\end{cases}
\end{equation*}
The corresponding normalized $6$-axes $\beta\Omega$-sequence equals\\
$\lr{1,2,3,4,-2,-1,5,6,6,1,-5,-3,2,1,-5,-3,2,1,-5,-6,-1,-2,-3,-4,-4,2,3,5,1,2,3,5,\ldots}.$\\
To obtain de $\beta,\Omega$ sequence, we replace $\pm 1,\pm 2,\pm 5$ with axis $\{\pm 1\}$ and $\pm 3,\pm 4,\pm 6$ with axis $\{\pm2\}$. 
Notice $T(-x)=-T(x)$.
We did not find an isometry $\s$ on the alphabet $\Sigma_6$, such that $T\big(\s(x)\big)=\s\big(T(x)\big)$ for all $x\in \Sigma_6$.

\subsection{Gray curve}\label{sc:graycurve}
The following example is distinctive: a fractal from which the $k$-curve exists in $k$-dimensional space and not in fewer dimensions.
We call the resulting curve the \emph{Gray curve} because the coordinates of the vertices of the curve form the \emph{binary reflected Gray code}.

\noindent\begin{minipage}{8cm}
	In \cref{fig:brGc}, we construct the binary reflected Gray code in dimension $d$, an ordered set of the binary vertices of the unit cube.
	
	First, we take the vertices in dimension $d-1$ and suffix each vertex with a $0$ coordinate. 
	Second, we reflect the same $d-1$-dimensional vertices, i.e., place them in reverse order, and suffix them with a coordinate $1$. 
	Finally, we join these two sets of vertices with the last coordinate of $0$ and $1$ in that order, respectively.

	Besides the binary reflected Gray code, different other Gray codes on the unit cube are possible. Nevertheless, we use the term ``Gray code'' for the binary reflected Gray code, and ``Gray curve'' for the curve of that code.
\end{minipage}\hfill
\begin{minipage}[c][8.0cm][t]{8cm}
	\centering
	\begin{figure}[H]
		\centering
		\includegraphics[scale=1.3]{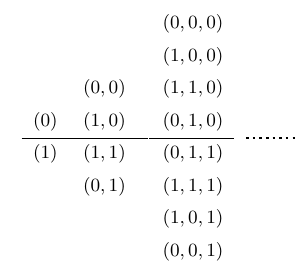}
		\caption{\small binary reflected Gray code} 
		\label{fig:brGc}
	\end{figure}
\end{minipage}

Define the \emph{Gray function} as $g_d:\{1,2,\ldots,2^d-1\}\to \{\pm 1,\pm 2,\ldots,\pm d\}$ such that if the subsequent vertices $v(n-1)$ and $v(n)$ in the Gray code differ $1$ in coordinate $k$, then $g_d(n)=k$, and if they differ $-1$, then $g_d(n)=-k$.\footnote{~$g_d$ Knuth  called a ``delta'' function \cite[p.~293]{Knuth.4.1}.}
This gives the sequence $\Lr{g_d(1),g_d(2),\ldots,g_d(2^d-1)}=\lr{1,2,-1,3,\ldots,-1}$ and the next definition as a consequence.

\begin{definition}\label{def:grayseq}
	The (binary reflected) \textbf{Gray sequence} $G$ is an infinite-dimensional sequence in $\Sigma^{\N}$, where $\Sigma= \Z\backslash\{0\}$.
	Its approximants $G(d)$ we define as $G(0)=\lr{}$ and for $d>0$,\\
	$G(d)=\Lr{g_d(1),g_d(2),\ldots,g_d(2^d-1)}= \LR{G(d-1),d,-\Rev\big( G(d-1)\big)}$, where $-\Rev$ is the inverse.
\end{definition}

\begin{figure}[H]
	\centering
	\includegraphics[scale=1]{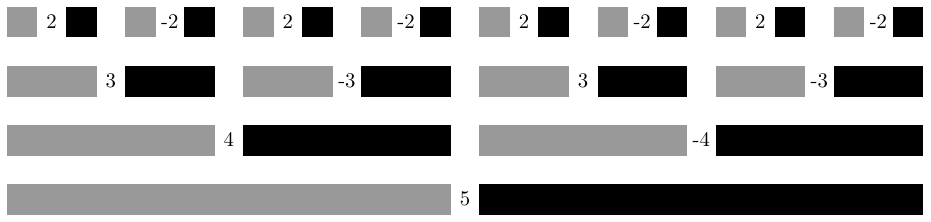}
	\caption{\small Visual of the Gray sequence. 
		A grey block represents the sub-sequence $G(k-1)$, followed by $\lr{\pm k}$, which is followed by a black block representing $-\Rev \big(G(d-1)\big)$, the inverse of a grey block.
		Each block is equal to the corresponding part of one of its upper lines.
		Notice that we have 
		$G(k-1)\supset G(k-2)\supset\cdots\supset G(2)\supset G(1)$ and 
		$-\Rev(G(k-1))\supset-\Rev(G(k-2))\supset\cdots\supset-\Rev(G(2))\supset-\Rev(G(1))$ for each $1<k$.}
	\label{fig:graysqblck}
\end{figure}

\noindent Note that 
$-\Rev\big(G(d)\big)=-\Rev\Lr{G(d-1),d,-\mathcal{R}\big( G(d-1)\big)} =\Lr{G(d-1),-d,-\Rev\big( G(d-1)\big)}$; therefore, 
$-\Rev\big(G(d)\big)= \big[1,2,\ldots,d-1,-d\big]G(d)$
as shown in \cref{fig:graysqblck}.

This Gray sequence appears under \seqnum{A164677} in  \cite{OEIS}\label{cite:OEIS3}.
It is normalized and starts with 
$\lr{1,2,-1,3,1,-2,-1,4,1,2,-1,-3,1,-2,-1,5,1,2,-1,3,1,-2,-1,-4,1,2,-1,-3,1\ldots}$.\\
Sloane \seqnum{A164677} observed that the Gray sequence is the paper-folding sequence 
$\m{Fold}(1,2,3,4,\ldots)$, mentioned in Exercise 15 in \cite[p.~203]{allshal}\label{cite:allshall1}.
This folding map is defined iteratively by \\
$\m{Fold}(x_1,\ldots,x_{n+1})=\big\la \m{Fold}(x_1,\ldots,x_n), x_{n+1}, -\mathcal{R}\big(\m{Fold}(x_1,\ldots,x_n) \big)\big\ra$\label{Fold} and $\m{Fold}(x_1)=\lr{x_1}$, similar to our definition of the Gray sequence.
We notice that the absolute value of the Gray sequence is the \emph{ruler function} in \seqnum{A001511} \cite{OEIS}.

There exist two substitutions that generate the Gray sequence: $T_1$ is \emph{uniform} (of length 2), c.f.~\cite{allshal}, that is, $\|T_1\lr{x}\|=\|T_1\lr{y}\|$ for $x,y\in \Sigma$, and $T_2$ is non-uniform.
\begin{equation*}
	\begin{cases}
		T_1(x)=\Lr{1,x+\sgn(x)}\text{ for } |x|=1,\footnotemark\\
		T_1(x)=\Lr{-1,x+\sgn(x)}\text{ for } |x|\not=1 \\
	\end{cases}\footnotetext{~This is the first of a series of uniform substitutions defined for $n>1$ by $T_n(x)=\Lr{G(n),x+n*\sgn(x)}$ for $|x|=1\text{ and } T_n(x)=\Lr{\Rev(G(n)),x+n*\sgn(x)}\text{ for }|x|\not=1$} 
	\begin{cases}
		T_2(1)=\lr{1,2,-1}\text{; } T_2(-1)=-\Rev\big(T_2(1)\big),\\
		T_2(x)=\Lr{x+\sgn(x)}\text{ for } |x|\not=1\\
	\end{cases}
\end{equation*}

\begin{definition}\label{def:graycrv}
	The (binary reflected) \textbf{Gray curve} $G$ is the curve on $\Z^{\N}$, which has the Gray sequence as description and the Gray code (with subsequent vertices connected) as a graph; $G(d)$, the $d^{\text{th}}$ approximant, lives on $\Z^d$.
\end{definition}

\begin{figure}[H]
	\centering
	\includegraphics[scale=0.8]{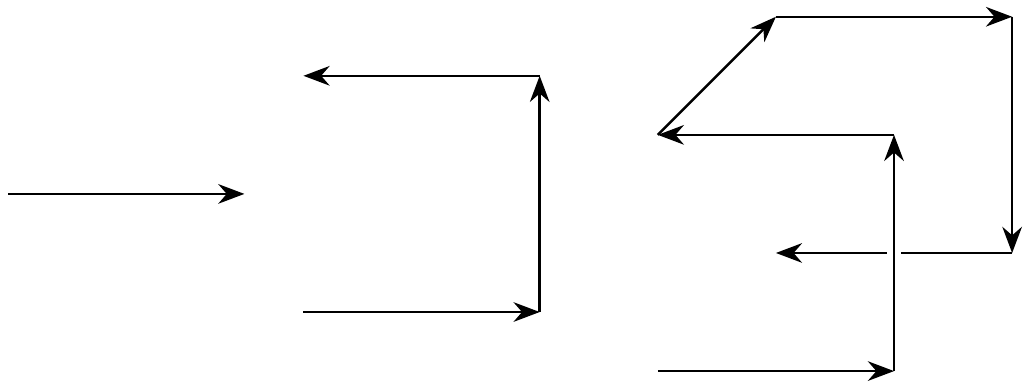}
	\caption{\small First three approximants of the Gray curve. The $3$-curve resembles a paperclip.} 
	\label{fig:graycrv123}
\end{figure}

Notice that $G(d)$ is a Hamiltonian path on the unit cube $C_d$, with the origin as the entry and the last vertex of the Gray code, $(0,0,\ldots,0,1)$, as the exit. 
Therefore, adding orientation to $G(d)$ transforms the Hamiltonian path into a Hamiltonian cycle.
\Cref{fig:graycrv123} shows the first few approximants, where the association with ``paper folding'' is evident.

\begin{observation}
	For $k=1,2,\ldots$, any set of $2^k$ subsequent edges in a Gray curve spans a $(k+1)$-dimensional unit cube $C_{k+1}$.
\end{observation}
\begin{proof}
	For no dimension $d$, there is a vertex in the Gray code outside the unit cube $C_d$ because all the vertices are in $L_\infty$-distance $1$ from the origin and have non-negative coordinates.	
	The number of vertices in the unit cube $C_d$ is $2^d$, all traced by the Gray curve $G(d)$.
	Hence, the number of edges in that path that trace each of these vertices only once is $2^d-1$.	
	As we observe from \cref{fig:graysqblck}, there are different sub-curves $H(j)\subset G(d)$ for $0\le j<d$ that are isometric with $G(j)$ because each gray block is a $G(j)$, and a black block is a $-\Rev\big(G(j)\big)$.
	
	Let $A=\lr{a_1,a_2,\ldots,a_{2^k}}$, of length $||A||=2^k$, be a (consecutive) sub-sequence $A\subseteq G(d)$, with $a\in A$ such that $|a|=\max\{|a_i|;1=1,2,\ldots,2^k\}$.
	Because $||A||=2^k>2^k-1$, it follows that $A\not\subseteq H(k)$, where $H(k)\cong G(k)$ (isometric); thus,$|a|>k$.
	
	Therefore, we have $A=\lr{a_1,a_2,\ldots,a_m,a=a_{m+1},a_{m+2},\ldots,a_{m+n+1}=a_{2^k}}$ with $0\le m<2^k$ and $m+n+1=2^k$;thus, $m\not=n$. 
	Thus, either $m<n$, in which case $2^k=m+n+1<2n+1$ and $2^{k-1}\le n$, or $n<m$ and $2^{k-1}\le m$.
	In the first case, $\big(H(k-1),\lr{k}\big)\subseteq \lr{a_{m+2},\ldots,a_{m+n+1}}$, and in the second case 
	$\big(\lr{k},(H(k-1)\big)\subseteq \lr{a_1,a_2,\ldots,a_m}$.
	In both cases, $A\setminus\{a\}$ counts $k$ directions, and hence the number of directions in $A$ equals $k+1$.
\end{proof}

If we consider $k=1$, then every two subsequent edges in a Gray curve are mutually orthogonal.

\begin{definition}
	A curve is ${n}$\textbf{-hyper-orthogonal} if for $k=1,2,\ldots,n$ any set of $2^k$ subsequent edges in the curve span a $(k+1)$-dimensional unit cube $C_{k+1}$.	
\end{definition}

Notice that if a curve is $n$-hyper-orthogonal, so are its isometric images.
Clearly, the Gray curve is $k$-hyper-orthogonal for $k>0$, and its approximant $G(d$) is $k$-hyper-orthogonal for $k\le d-1$.

We say that a curve in $\R^d$ is {hyper-orthogonal} if the curve is $(d-2)$-hyper-orthogonal. 
In three dimensions, this implies that a curve is hyper-orthogonal if and only if all subsequent edges are orthogonal to each other.

We can extend $G(d)$ with additional edges preceding the entry and succeeding the exit without losing the $(d-1)$-hyper-orthogonality by adding an edge $\lr{d}$ before and after the curve. 
We can build a chain of $G(d)$s, mutually connected by the edge $\lr{d}$, which is still $(d-1)$-hyper-orthogonal.

We used hyper-orthogonality \cite{BH} to construct Hilbert curves with excellent properties, as discussed in the following section.

\subsection{Dekking's Gosper-type curve} 
\begin{figure}[H]
	\centering
	\includegraphics[scale=0.35]{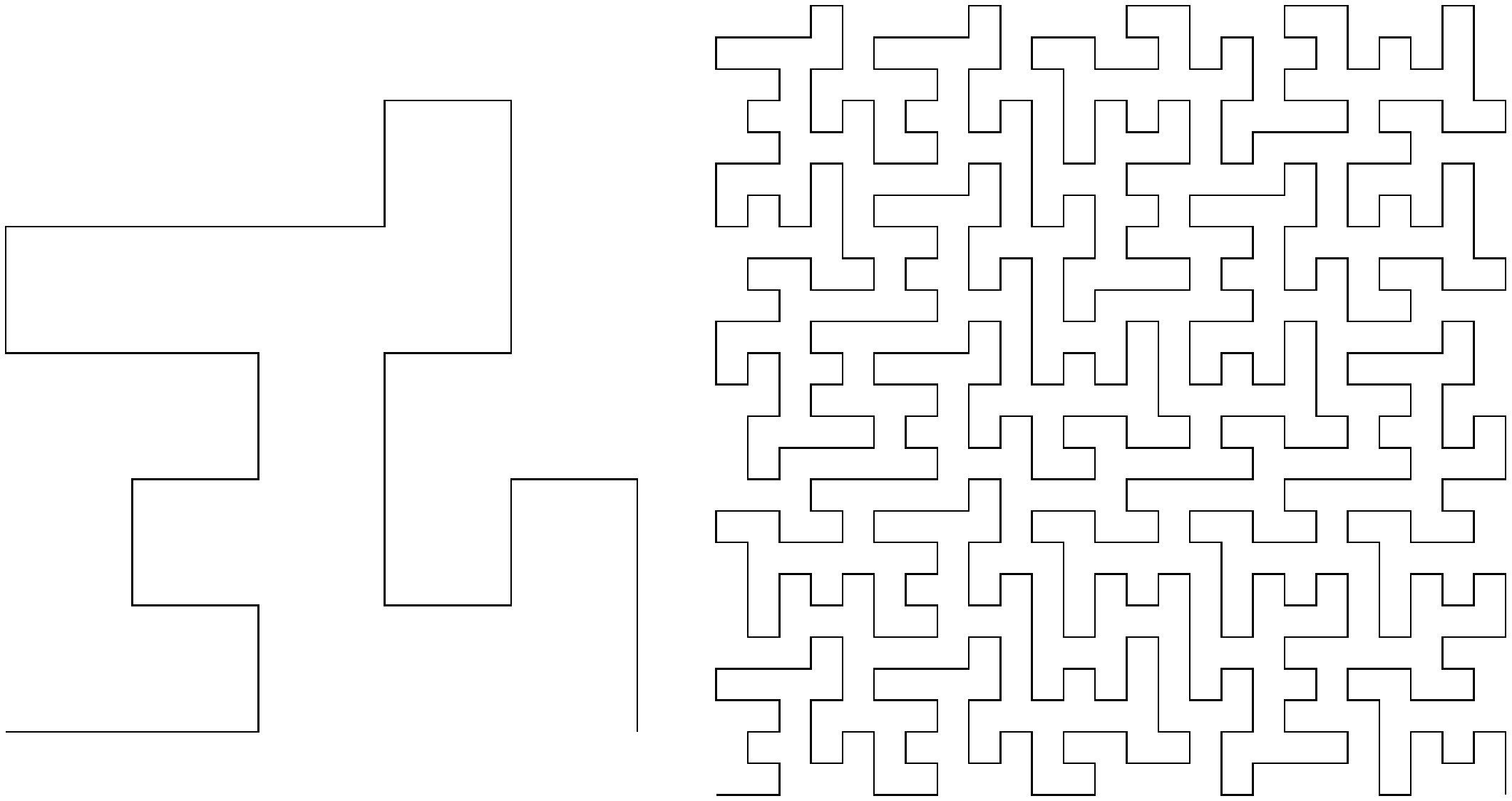}
	\caption{\small The $1$ and $2$-curve of Dekking's Gosper-type curve. See the last two drawings on \cref{fig:Dekking_flowsnake} (p.~\pageref{fig:Dekking_flowsnake}) in \cref{sub:Dkkng Flwsnk} for an artistic impression of the second and third approximant.} 
	\label{fig:DkFlsnk1}
\end{figure}

This curve was described by Dekking \cite{Dekk0} in $1982$; McKenna \cite{McK} found it in $1978$, called it ``E-curve'', but only published it in $1994$.
Dekking gave an extensive treatment for the method we apply here, be it in another notation.
We have the following translation table for the items he used in his example (4.9).
\[\begin{matrix}
	s_{00}&\mapsto&\lr{1}& s_{10}&\mapsto&\lr{-3}\\
	s_{01}&\mapsto&\lr{4}& s_{11}&\mapsto&\lr{2}\\
	s_{02}&\mapsto&\lr{-1}& s_{12}&\mapsto&\lr{3}\\
	s_{03}&\mapsto&\lr{-4}& s_{13}&\mapsto&\lr{-2}
\end{matrix}\]
Here we map $\lr{1}\mapsto(1,0)\mapsfrom\lr{-3}$, whereas $\lr{4}\mapsto(0,1)\mapsfrom\lr{2}$. 
Hence, the isometries used are 
\mbox{$\rho=\begin{bmatrix} 1 & 4 & -3 & 2\\ -3 & 2 & 1 & 4\\ \end{bmatrix}=[-3,4,-1,2]$,} which swaps the columns in the matrix above, and geometrically is the identity, and 
$\s=\begin{bmatrix}1 & 4 & -3 & 2\\ 4&-1&2&3\\\end{bmatrix}=[4,3,-2,-1]$, which rotates each column one step downward (in the matrix above), and geometrically is the rotation over $\pi/2$.
Dekking defined the substitution that we derive using our toolkit.
\begin{theorem}
	The normalized sequence for Dekking's Gosper-type curve is\\
	$\lr{1,1,2,-1,2,1,2,-1,-1,2,1,1,1,2,1,-2,-2,-1,-2,-2,1,2,1,-2,-2,1,1,2,-1,2,\ldots}.$\\
	This sequence is obtained by $\lr{1}$ as the start and the substitution \\
	$S(\iota)=(\iota,\iota,\mu\tau_y,-\tau,\mu,\iota,\mu\tau_y,-\tau,-\iota,\mu\tau_y,\iota,\iota,\tau,\mu,\tau,-\mu,-\mu,-\tau,-\mu,-\mu\tau_y,\tau,\mu,\iota,-\mu\tau_y,-\mu\tau_y)$\\
	where $\mu=[2,-1]$ is the rotation over $\pi/2$, $\tau_y=[1,-2]$ is the vertical reflection, $\iota=[1,2]$ the identity and $-\iota=[-1,-2]$ the inverse. The first two approximants are shown in \cref{fig:DkFlsnk1}.
\end{theorem}
\begin{proof}
In \cref{fig:DkFlsnk}, we show the derivation of the sequence and the substitution for the $1$-curve of this Gosper-type curve.
In the drawing on the left, we observe that the primary edge below, $\lr{1}$, covers the $5\times 5$ square on its left side.
Therefore, we determine the mapping between squares and edges in the center image by starting with the $(13)$ dark gray colored squares with a unique edge. 
Each square contains the direction of its unique edge, $\pm 1,\pm 4$ if the square is at the left of the directed edge, or $\pm 2,\pm 3$ otherwise.
Subsequently, in the middle drawing, the $(4)$ edges in a unique, light gray colored square we map to the corresponding squares.
Finally, we similarly treat the ($8$) remaining unattached edges and white squares, as shown in the drawing rightward to \cref{fig:DkFlsnk}.
This process leads to the substitution\\ 
$S\lr{1}=\lr{1,1,2,3,4,1,2,3,-1,2,1,1,-3,4, -3,-4,-4,3,-4,-2,-3,4,1,-2,-2}$.
\begin{figure}[H]
	\centering
	\includegraphics[scale=0.9]{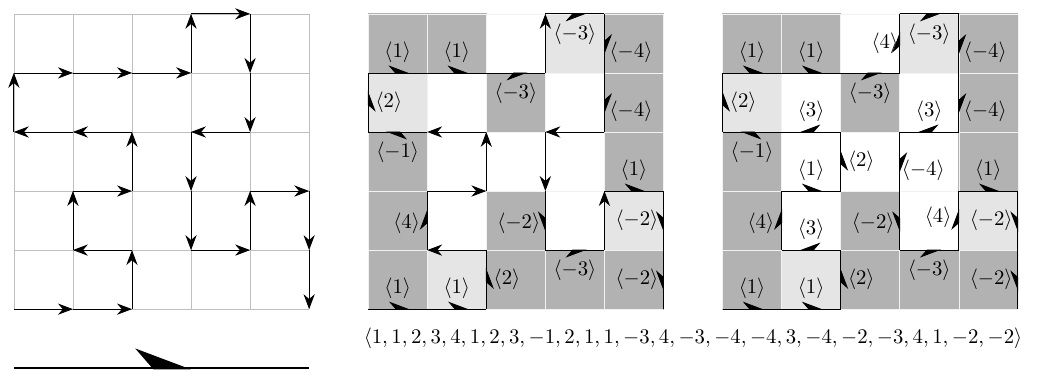}
	\caption{\small Construction of sequence and substitution of the $1$-curve of Dekking's Gosper-type curve. 	See \cref{fig:Dekking_flowsnake} (p.~\pageref{fig:Dekking_flowsnake}) in \cref{sub:Dkkng Flwsnk} for an artistic impression of the second and third approximants.} 
	\label{fig:DkFlsnk}
\end{figure}

If we split Dekking's $\rho$ by the two reflections $\tau_y'=[-3,-4,-1,-2]$ and $\tau_x'=[3,4,1,2]$, and replace his $\s$ by the rotation $\mu'=[4,3,-2,-1]$, we see that $\mu'\tau_y'\lr{1}=\lr{2}$, $-\tau_y'\lr{1}=\lr{3}$, and $\mu'\lr{1}=\lr{4}$. 
As for the substitution $S$, we have $S(\varphi)=\varphi S(\iota)$; for all isometries $\varphi$, and we get\\
{\small$ S(\iota)=(\iota,\iota,\mu'\tau_y',-\tau_y',\mu',\iota,\mu'\tau_y',-\tau_y',-\iota,\mu'\tau_y',\iota,\iota,\tau_y',\mu',\tau_y',-\mu',-\mu',-\tau_y',-\mu',-\mu'\tau_y',\tau_y',\mu',\iota,-\mu'\tau_y',-\mu'\tau_y')$.}
\begin{figure}[H]
	\centering
	\includegraphics[scale=0.9]{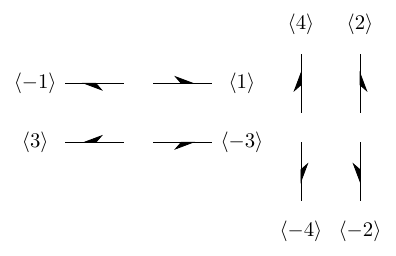}\qquad\qquad
	\caption{\small The arrows left from which we learn that $\mu'=[4,3,-2,-1]$ is the rotation over $\pi/2$ and \mbox{$\tau_y'=[-3,-4,-1,-2]$ and $\tau_x'=[3,4,1,2]$} are the vertical and horizontal reflection in the $4$-axes plane.} 
	\label{fig:Dkarrws}
\end{figure}
We project $-3\mapsto 1$ and $4\mapsto 2$, and if we apply that, we find that replacing $\mu'$ by $\mu=[2,-1]$, the rotation over $\pi/2$ in the plane, and $\tau_y', \tau_x'$ by $\tau_y=[1,-2], \tau_x=[-1,2]$, the vertical and horizontal reflection.
The substitution and the resulting sequence then become\\
{\small $S(\iota)=(\iota,\iota,\mu\tau_y,-\tau_y,\mu,\iota,\mu\tau_y,-\tau_y,-\iota,\mu\tau_y,\iota,\iota,\tau_y,\mu,\tau_y,-\mu,-\mu,-\tau_y,-\mu,-\mu\tau_y,\tau_y,\mu,\iota,-\mu\tau_y,-\mu\tau_y),$}\\
$\lr{1,1,2,-1,2,1,2,-1,-1,2,1,1,1,2,1,-2,-2,-1,-2,-2,1,2,1,-2,-2,1,1,2,-1,2,\ldots}.$\\
As Dekking's Gosper-type curve is normalized from the beginning,
we submitted this sequence to the \href{https://oeis.org}{OEIS} as \seqnum{A356112}.

\end{proof}

\subsection{Arndt's Peano curve}
Arndt \cite{Arndt1} investigated all the fractal space-filling curves that can be constructed using a Lindenmayer system \cite{AL} with only one variable, using a result from Dekking \cite{Dekk1}.
This section discusses his case R9--1, the Peano curve on the square grid \cite{Sagan}. As Arndt showed, we can also draw this curve on the \emph{truncated} square grid.

We observe that the substitution is relatively simple because it only uses the minimal rotation $\mu=[2,-1]$ over $\pi/2$, i.e.,
$T(\iota)=(\iota,\mu,\iota,-\mu,-\iota,-\mu,\iota,\mu,\iota)$. 
Note that the last four transformations are equal to the first four in reverse order, which displays the folding characteristic of the curve.
But Dekking \cite[Example (4.1)]{Dekk0} already pointed out in his first example that the substitution $T_D$ given by  $T_D(\iota)=(\iota,\mu,\iota)$ and $T_D(\mu)=(-\mu,-\iota,-\mu)$\footnote{~Notice here we do not have $T_D(\mu)=\mu \circ T_D(\iota)$}, is enough.
The normalized sequence is\\
$\lr{1,2,1,-2,-1,-2,1,2,1,2,-1,2,1,-2,1,2,-1,2,1,2,1,-2,-1,-2,1,2,1,-2,1,-2,\ldots}$\\
which has as simple substitution $T$ with $T(-x)=-T(x)$
\begin{equation*}
	T=\begin{cases}
	1 &\to\lr{ 1,2,1,-2,-1,-2,1,2,1}\\
	2 &\to\lr{ 2,-1,2,1,-2,1,2,-1,2} \\
	\end{cases},
\text{ or even simpler: }
	T_D=\begin{cases}
	1 &\to\lr{ 1,2,1}\\
	2 &\to\lr{-2,-1,-2} \\
\end{cases}
\end{equation*}

\begin{figure}[H]
	\centering
	\includegraphics[scale=0.8]{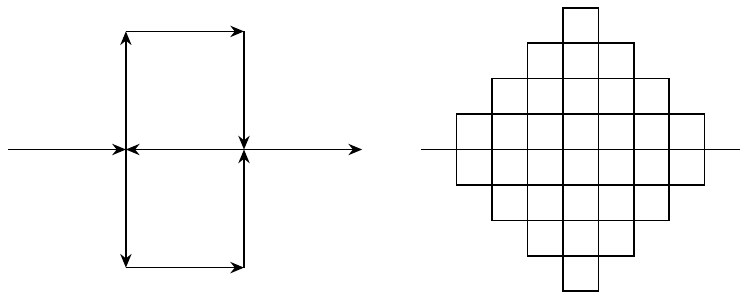}
	\caption{\small First two approximants of Arndt's R9--1, the Peano curve on the square grid.} 
	\label{fig:R9-1}
\end{figure}

This curve is peculiar because the curve exhibits a higher degree of space-filling. 
Usually, similar to the Hilbert curve, a space-filler visits each vertex of the grid the curve lives on only once, and some edges of that grid the curve never visits; these are called space-filling curves and are, in fact, \emph{vertex-covering curves}.
However, this Peano curve visits each \emph{edge} of the square grid once and consequently each vertex exactly twice, and these curves are called \emph{edge-covering curves}.

The author \cite{Arndt1} used the R9--1 square grid curve, \cref{fig:R9-1}, to extend it to the truncated square grid, as we depict on the right-hand side of \cref{fig:hc_trsq_grd} (page \pageref{fig:hc_trsq_grd}).
This extension is relatively simple by considering the right-hand side drawing of \cref{fig:hc_trsq_grd_2} (page \pageref{fig:hc_trsq_grd_2}).
Here, we can observe that a direction in the square grid generates \emph{a pair} of subsequent directions in the truncated square grid, such as $\lr{1}\to \lr{1,2}$, if the first edge was part of a pair $\lr{1,2}$, and $\lr{1}\to \lr{1,-4}$, if the first edge was part of a pair $\lr{1,-2}$.

\begin{figure}[H]
	\centering
	\includegraphics[scale=1]{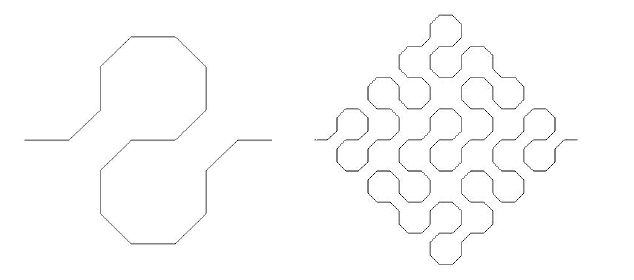}
	\includegraphics[scale=0.9]{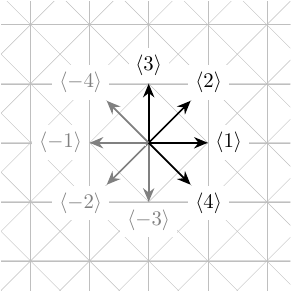}
	\caption{\small First two approximants of R9-1 on the truncated square grid, with its generators.} 
	\label{fig:TSqr_R9-1}
\end{figure}

Therefore, to get the fractal on the truncated square grid, we split the sequence for the square grid into overlapping pairs, such as $\lr{1,2},\lr{2,1},\lr{1,-2},\ldots$, and then apply the following substitution from sequences in the alphabet
$\Sigma_2\times \Sigma_2$ to sequences in $\Sigma_4\times \Sigma_4$.
\begin{equation*}
	T'=\begin{cases}
		\lr{1,2}&\to\;\lr{1,2}\\
		\lr{1,-2}&\to\;\lr{1,4}\\
		\lr{2,1}&\to\;\lr{3,2} \\
		\lr{2,-1}&\to\;\lr{3,-4} \\
	\end{cases}
	\quad\text{ together with }T'\lr{-x,-y} = -T'\lr{x,y}
\end{equation*}
 and after concatenating all pairs from $\Sigma_4\times \Sigma_4$, we get the sequence for the truncated square grid\\
$\lr{1,2,3,2, 1,4,-3,-2, -1,-2,-3,4,1,2,3,2,1,2,3,-4,-1,-4,3,2,1,4,-3,4,1,2,3,\ldots}$. \\
This sequence has the drawings in \cref{fig:TSqr_R9-1} as the first two approximants, which are similar to those of Arndt \cite[sec.~4.1.2]{Arndt1}.
We did not find any simpler substitution for this sequence.

\subsection{Ventrella's Box 4}\label{sub:box4}

\begin{figure}[H]
	\centering
	\includegraphics[scale=1.0]{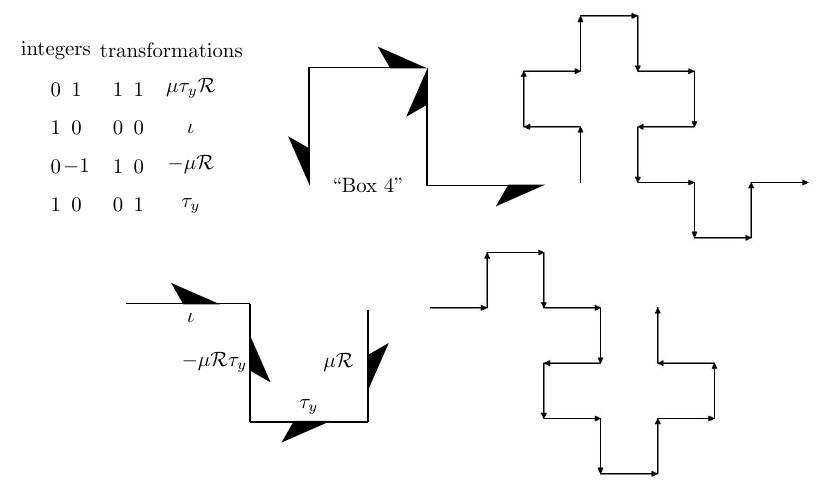}
	\caption{\small Top row: Ventrella's notation and his flagged $1$-curve; bottom row: our adaptation.}
	\label{fig:ventrll1crvs}
\end{figure}

Ventrella \cite{VentrellaTree} studied a fractal called ``Box 4,'' clearly indicated by his original flags, which conform to the center drawing of the first row of \cref{fig:ventrll1crvs}, where the $2$-curve of the fractal is at the right-hand side.
We added the first two columns, ``integers'' and ``transformations,'' using his notation, plus the column using our notation, with $\Rev,\tau_y,\text{ and }\mu$, as in \cref{fig:ventrllflags,fig:ventrllflags2,fig:otherflags}.
Using our centered flags with their corresponding transformations in the second row, we transformed his curves to be normalized and extending (c.f.~page \pageref{df:extending}).

\begin{figure}[H]
	\centering
	\includegraphics[scale=1.05]{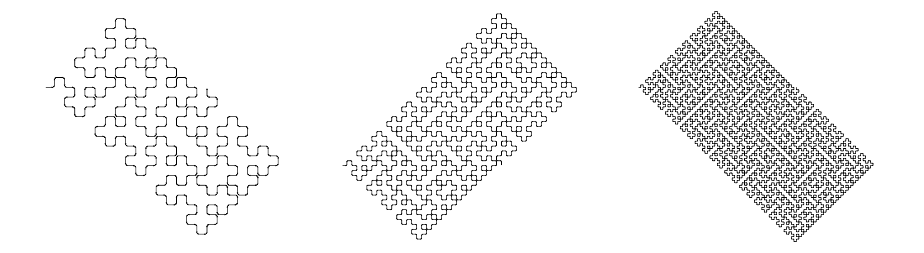}
	\caption{\small $k$-curves for the Box 4-fractal with $k=3,4,5$, all with rounded corners.}
	\label{fig:box4crvs}
\end{figure}

\begin{theorem}
	With $T(\iota)=\big(\iota,\tau_{-d}\Rev,\tau_y,\mu\Rev\big)$ we have $S_{k}=\tau_y^k T(S_{k-1})$, where $\iota=[1,2]$ is the identity, $\tau_y=[1,-2]$ the vertical reflection, $\tau_{-d}=[-2,-1]$ the anti-diagonal reflection, and $\mu=[2,-1]$ the minimal rotation over $\pi/2$.~\footnote{~In \cref{sc:Cayley}, we describe the group of isometries of the square grid with the signed permutations.}
\end{theorem}
\begin{proof}
We can use the $2$-curve (right-hand side of the first row in \cref{fig:ventrll1crvs}) to determine the four transformations of the $1$-curve with which we construct the $2$-curve. First, we normalize Ventrella's approximants by applying reverse $\Rev$ and setting $\lr{1,2,1,-2}$ for the $1$-curve and\\ 	
$\lr{1,2,1,-2,1,-2,-1,-2,1,-2,1,2,1,2,-1,2}$ for the $2$-curve (right-hand side of the second row in \cref{fig:ventrll1crvs}). We partition the $2$-curve into discrete curves of length four and determine their isometric images of the $1$-curve, the first being $\iota$.
The other images are 
$\lr{1,-2,-1,-2}=\tau_{-d}\Rev\lr{1,2,1,-2}$, then $\lr{1,-2,1,-2}=\tau_y\lr{1,2,1,-2}$, and finally $\lr{1,2,-1,2}=\mu\Rev\lr{1,2,1,-2}$, which brings the substitution of the isometries into $T(\iota)=\big(\iota,\tau_{-d}\Rev,\tau_y,\mu\Rev\big)$.

However, we must overcome an obstacle that did not occur in Ventrella's approach.
We not only want our sequences to be normalized but also to be``extending'' (p.~\pageref{df:extending}).
If we apply our substitution $T(\iota)=\big(\iota,\tau_{-d}\Rev,\tau_y,\mu\Rev\big)$ to the $0$-curve, which is edge $\lr{1}$, we obtain for a $1$-curve $\lr{1,-2,1,2}$, which is not the start of the $2$-curve. 
Fortunately, we can resolve this by applying an additional $\tau_y$ after the substitution to get a $k$-curve, where $k$ is odd.
Therefore, we get $S_{k}=\tau_y^k T(S_{k-1})$ because $\tau_y^2=\iota$.
\Cref{fig:box4crvs} shows a few approximations of the normalized and extending Box4-fractal.
If we apply this substitution to the start edge $\lr{1}$, we get\\
$\lr{1,2,1,-2,1,-2,-1,-2,1,-2,1,2,1,2,-1,2,1,-2,1,2,1,2,-1,2,-1,-2,-1,2,\ldots}.$
\end{proof}

\subsection{Ventrella's V1 Dragon}
\begin{figure}[H]
	\centering
	\includegraphics[scale=1.2]{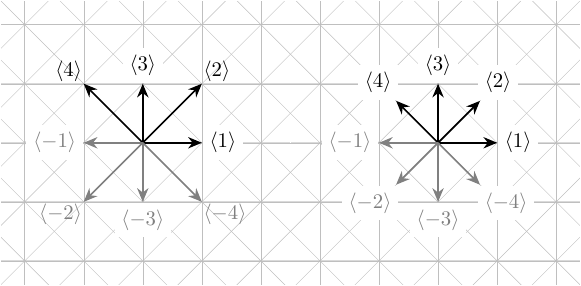}
	\caption{\small Square-diagonal grid and $8^\text{th}$-roots grid.}
	\label{fig:sqrdiag8roots}
\end{figure}

This example uses the square-diagonal grid, which is peculiar because not all directions have the same lengths; refer to the grid on the left-hand side in \cref{fig:sqrdiag8roots}. On comparing the two grids, the $8^\text{th}$ roots of unity span the right one, where all directions have equal lengths.
However, there is a significant difference between the two grids. 
The minimal rotation in the $8^\text{th}$ roots grid is $\mu=[2,3,4,-1]$, since all generators have equal length.
However, the minimal rotation in the square diagonal grid is $\mu=[3,4,-1,-2]$ because we have two sets of generators of different lengths.
Thus, we could ask whether we would need the square diagonal grid. 
The essential difference between the two grids is that the vertices of the square diagonal grid are equal to $\Z^2$, whereas the vertices of the $8^\text{th}$ roots grid are dense in $\R^2$.
In this example, we will show how grids on the square-diagonal grid can be constructed with all features of the $8^\text{th}$ roots grid.
The minimal rotation in $8^\text{th}$ roots grid is over $\pi/4$, and is given by $\mu=[2,3,4,-1]$, whereas the vertical reflection is $\tau_y=[1,-4,-3,-2]$.
As the directions are mutually dependent, the two signed permutations do not generate the entire hyper-octahedral group of four dimensions. However, they generate the symmetry group of the octagon, i.e., the dihedral group $D_8$.

\begin{figure}[H]
	\centering
	\includegraphics[scale=0.7]{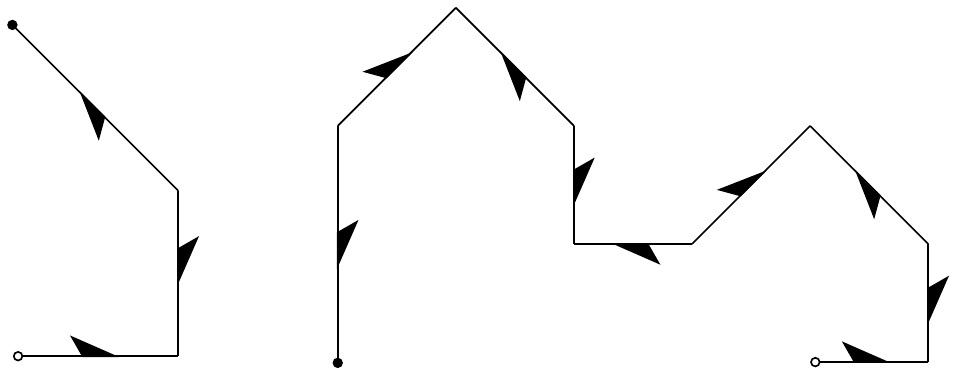}
	\caption{\small $1$- and $2$-curves of Ventrella's V1 Dragon.}
	\label{fig:V1drgn}
	\includegraphics[scale=0.7]{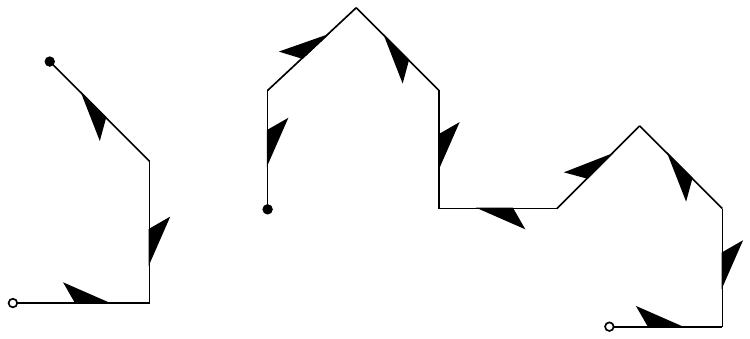}
	\caption{\small Same $1$- and $2$-curves on the $8^{\text{th}}$-root grid.}
	\label{fig:V1drgnrt8}
\end{figure}

We consider another example of Ventrella \cite{VentrellaTree}, called the ``V1 Dragon,'' that ``lives'' on the square-diagonal grid.
We slightly altered his sample to make it normalized and extending. In \cref{fig:V1drgn}, we observe its $1$- en $2$-curves.
Similar to the previous \cref{sub:box4}, we can ``read'' the transformations involved from the first picture by using only reverse $\Rev$ and rotation $\mu$ (and the identity $\iota$). Therefore,
$T(\iota)=\big(\iota, \Rev\mu^2, \sqrt{2}\ast\mu^3 \big)$.

\begin{figure}[H]
	\centering
	\includegraphics[scale=0.5]{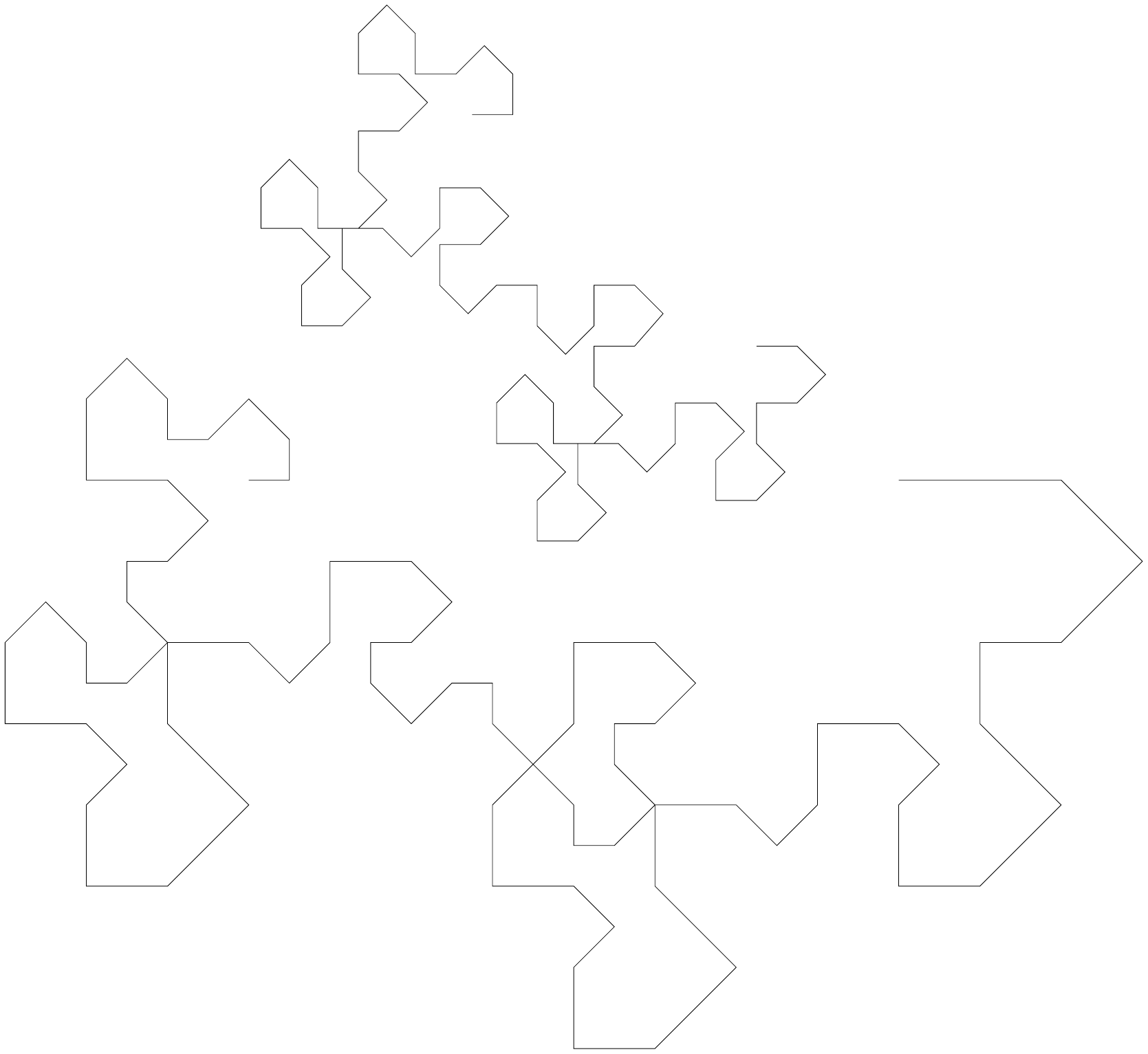}
	\caption{\small The V1-Dragon $4$-curves on the $8^{\text{th}}$-roots and the square-diagonal grids.}
	\label{fig:V1drgns}
\end{figure}

Here, we notice an important difference with previous fractals, which only had edges of length one. Apart from the transformations $\mu$ and $\Rev$, we multiply the last length with $\sqrt{2}$.
Note that the $2$-curve (second picture in \cref{fig:V1drgn}) has last edge with length $2=(\sqrt{2})^2$.
If we draw the same sequence in the $8^{\text{th}}$-roots grid, we have almost the same fractal, but with edges of length one, as shown in \cref{fig:V1drgnrt8}.
As this grid is dense in $\R^2$, the geometric picture is less pleasing than the one in the square-diagonal grid, as can be seen in \cref{fig:V1drgns}.

As we cannot determine the length of a fractal sequence from the directions in its representation, we introduce a separate sequence of lengths using its \emph{length substitution}, which, completely independent from the \emph{directional substitution}, determines the geometrical fractal from both series.

In \cref{fig:V1drgns}, we see the $4$-curves of Ventrella's V1 dragon, with the upper one on the $8^{\text{th}}$-root grid and the lower one on the square-diagonal grid.
The lower one is larger because some of the edges have grown in length, and the vertices of this curve are on the lattice $\Z^2$. 
By contrast, the upper curve has edges that \emph{partially} overlap, which, to the best of our knowledge, is unseen in geometrical fractals. A fractal curve only shares vertices with itself, or edges, or neither of the two.

As we are more interested in fractals as number sequences than geometrical figures, we have the \emph{same sequence for directions} in both grids, and a length sequence in the case of the square-diagonal grid.
Therefore, we first split the substitution $T$ into two: $_dT$ for the directions and $_lT$ for the lengths, which lead to the sequences $_dS$ and $_lS$, respectively.
Thus, we get 
\begin{equation*} 
	T(\iota)=\big(\iota, \Rev\mu^2, \sqrt{2}\ast\mu^3 \big) \equiv
	\begin{cases}
		T_d(\iota)=\big(\iota, \Rev\mu^2, \mu^3 \big)\\
		T_l(\iota)=\big(\iota, \Rev, \sqrt{2} \big)
	\end{cases}
\end{equation*}
Starting with $S_x(0)=\lr{1}$ for both $x=d,l$, this leads to the sequences \\
$S_d=\lr{1,3,4,-2,-1,3,4,-2,-3,1,-4,-2,-1,-3,-4,-2,-1,3,4,-2,-3,1,-4,-2,\ldots}$ and\\
$S_l=\lr{1,1,\sqrt{2},\sqrt{2},1,1,\sqrt{2},\sqrt{2},2,2,\sqrt{2},\sqrt{2},1,1,\sqrt{2},\sqrt{2},1,1,\sqrt{2},\sqrt{2},2,2,\sqrt{2},\sqrt{2},2,2,\ldots}$. 
The latter can be simplified by taking the $\sqrt{2}$ logarithm, which gives\\ 
$\log_{\sqrt{2}}\left(S_l\right)=\lr{0,0,1,1,0,0,1,1,2,2,1,1,0,0,1,1,0,0,1,1,2,2,1,1,2,2,3,3,2,2,1,1,2,2\ldots}$,
which is the double \big(i.e., $\lr{x,x}$ versus $\lr{x}$\big) of $A062756$  \cite{OEIS}.

The sequence $S_d$ is not normalized because the numbers $3$ and $4$ are used before $2$, as can be observed from the first directions with different axes in our definition of V1 dragon, which are $1,3,4,-2$; check the $2$-curve in \cref{fig:V1drgn} or \ref{fig:V1drgnrt8}.
For this case, we have the normalizing permutation (c.f.~\cref{def:char_perm}), which happens to be $[1,-4,2,3]$ and brings the sequence back to:\\
$S_d'=\lr{1,2,3,4,-1,2,3,4,-2,1,-3,4,-1,-2,-3,4,-1,2,3,4,-2,1,-3,4,-2,1,-4,3\ldots}$.
Suppose we want this normalized sequence to represent the isometric image of the V1 dragon. In this case, we must adjust the numbering of the directions as the normalizing permutation indicates, being the analog of a base transformation in a vector space.
\begin{figure}[H]
	\centering
	\includegraphics[scale=1.2]{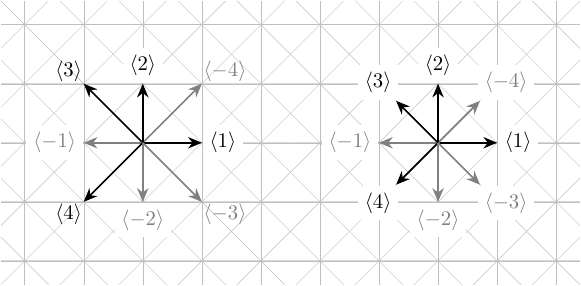}
	\caption{\small The two grids we number differently than in \cref{fig:sqrdiag8roots}.}
	\label{fig:sqrdiag8rootsb}
\end{figure}

Therefore, we changed the numbering of the directions, such that $1,3,4,-2$ becomes $1,2,3,4$, as in \cref{fig:sqrdiag8rootsb}, i.e., with $[1,3,4,-2]^{-1}=[1,-4,2,3]=\s$.
The transformations we derive from \cref{fig:V1drgn}, are the same; however, the minimal rotation we now represent by $\mu'=[-4,3,-1,-2]$, which is equal to 
$\mu'=\s\mu\s^{-1}$, where $\mu=[2,3,4,-1]$ is the minimal rotation in the original directions, and $\s$ is the ``base transformation.''
Therefore, the new substitutions become 
$T_d'(\iota)=\big(\iota, \Rev\mu'\,^2, \mu'\,^3 \big)$ and $T_l'=\,T_l$, and we get the same geometric pictures as \cref{fig:V1drgns}.

\section{Considerations and conclusions}
\subsection{Conclusions}
In this study, we have an alphabet that we use to describe the fractal sequence. This alphabet is our translation of the grid on which the fractal image exists.
We explicitly map the set of geometrical line-fractals to the set of signed integer sequences.
We provide a substitution, and a starter sequence from which the fractal grows, via its approximants, which are finite approximations of the limit fractal.

The advantage of our approach is threefold. 
First, we can order the set of normalized, signed, and integer sequences, which implies an ordering on the set of fractal images. 
Second, we can use the machinery of signed permutations as isometries of signed integer sequences. 
Notably, the ``reverse'' appears to be an essential anti-morphism. 
Finally, the ``mapping'' of a fractal image to a signed integer sequence makes it sufficiently simple to obtain the image from that sequence.

We describe our finding using seven examples with eight? sequences to illustrate the different peculiarities encountered when representing a fractal sequence as a signed integer sequence.

Finally, we set up an inventory of the fifteen fractal integer sequences, most of which are unknown in the On-Line Encyclopedia of Integer Sequences \cite{OEIS}.

\subsection{Considerations}
We recognize our preference for integer numbers, contrary to others who prefer natural numbers, letters, or other systems.
In our opinion, it is a matter of taste, and we like to work with signed permutations, while others work with L-systems or finite machines.

The fractal sequences we discuss here to illustrate the method of mapping a fractal to a signed integer sequence are superficial.
Therefore, the primary task should be to scan all publications describing fractals, convert them into integer sequences, and add them to the list we started in our study, which can build an ordered catalog of fractal sequences.
The fractals in the publications of Mandelbrot, Dekking, Arndt, and Ventrella we will describe first, not necessarily in that order.

Subsequent research could use the method described here to generate new fractal sequences comparable to what Arndt did with the Lindenmayer systems.

In further studies, emphasis should be given to (line) fractals in higher dimensions, i.e., above two, as most known fractals are on the cubic grid, like the Gray curves and their offspring, i.e., the Hilbert curves.

Finally, the fractals occurring in this study are line-fractals.
We are anxious to know how to represent fractals with higher dimensional structures, like planes or volumes.


\bibliographystyle{plainnat}

\appendix
\appendixpage
\addappheadtotoc
\section{Dihedral group \texorpdfstring{$D4$}{D4}}\label{sc:Cayley}

\begin{figure}[H]
	\centering
	\includegraphics[scale=0.9]{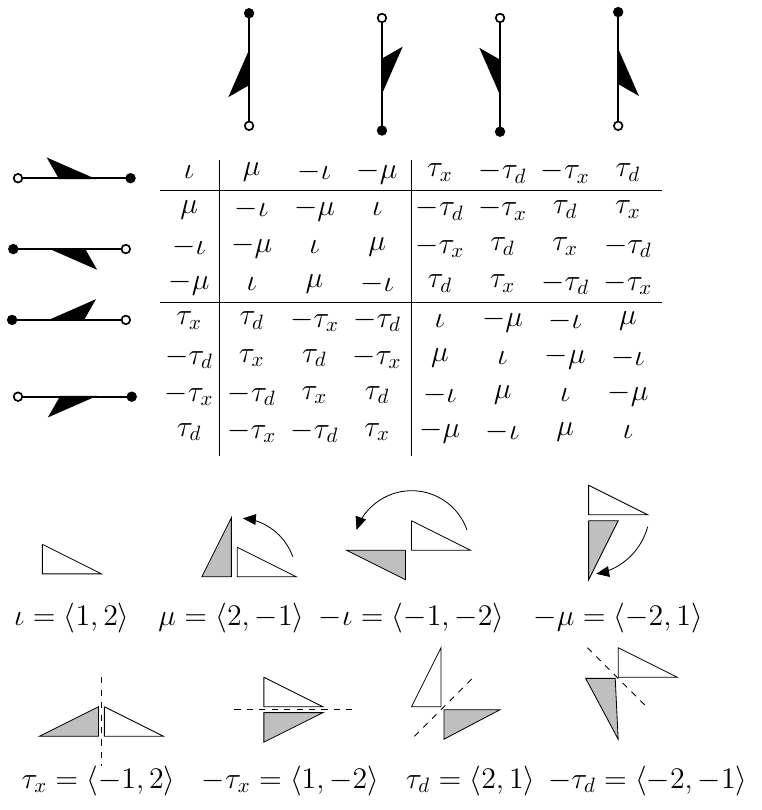}
	\caption{\small Cayley table of some transformations of the square grid with their pictorial description below and its results on flags. It indicates the reflections, such as $\tau_x$ because the $x$-coordinate changes, $\tau_d$ because the reflection is diagonal.}
	\label{fig:btOmIso} 
\end{figure}
\begin{observation}
	The group of signed permutations in dimension $d$ is the hyper-octahedral group of order $(2d)!!=2^d d!$ .
	In two dimensions, this group is generated, among others, using the minimal rotation $\mu=[2,-1]$ and the vertical reflection $\tau_y=[1,-2]$, which we use for the square grid throughout this study. 
\end{observation}

\section{Normalized Fractal Encyclopedia}\label{sc:NrmFrEnc}
\subsection{Description}
An (integer) sequence essentially determines its fractal, and because we cannot characterize an infinite sequence, we use a picture and need \emph{the determining alphabet} and its \emph{substitution}.
To depict the fractal, we need the generators of a grid and their relation with the alphabet. What follows is a template of how we store a fractal.

\begin{description}
\setlength{\itemsep}{0ex}
	\item[sequence:] We will provide approximately the first $20$ numbers of the sequence.
	\item[in \href{https://oeis.org}{OEIS}:] its occurrence (signed) in the \href{https://oeis.org}{OEIS}, plus the inspection date.
	\item[alphabet:] The alphabet of the fractal.
	\item[substitution:] The substitution of the fractal.
	\item[grid:] The grid on which the fractal is depicted.
	\item[generators:] The generators of the grid in the order of the positive elements of the alphabet.
\end{description}
In most cases, we provide the graphs of some initial $k$-curves.
In all cases, we use the following nomenclature
\begin{itemize}
\setlength{\itemsep}{0ex}
	\item The \emph{identity isometry} $[1,2,\ldots,d-1,d]$ we denote by $\iota$.
	\item With $\mu$ we represent the minimal rotation $[2,3,4,\ldots,d-1,d,-1]$.
	\item The notation $-\iota$ is shorthand for the isometry $[-1,-2,\ldots,-d]$.
	\item The \emph{reverse} $\Rev$, only defined on a finite sequence, by $\Rev\Lr{s(1),s(2),\ldots,s(n-1),s(n)}=\Lr{s(n),s(n-1),\ldots,s(2),s(1)}$, is the anti-morphism, such as an approximant of a fractal.
\end{itemize}

The \textbf{bold}number in a sequence is the first number that makes this sequence differ from, and larger than the previous one.

\subsection{Index of fractal sequences}
$\lr{\mathbf{1},1,2,-1,2,1,2,-1,-1,2,1,1,1,2,1,-2,-2,-1,-2,-2,\ldots},$\\ 
\indent c.f.~\ref{sub:Dkkng Flwsnk}, Dekking's Gosper-type curve on the square grid.

\noindent$\lr{1,1,2,\mathbf{3},4,1,2,3,-1,2,1,1,-3,4,-3,-4,-4,3,-4,-2,-3,4,1,-2,-2,1,1,2,3,4,\ldots}$,\\
\indent c.f.~\ref{sub:Dkkng Flwsnk2}, Dekking's Gosper-type curve on a $4$-axes grid.

\noindent$\lr{1,\mathbf{2},1,-2,1,-2,-1,-2,1,-2,1,2,1,2,-1,2,1,-2,1,2,\ldots},$\\ \indent c.f.~\ref{sub:Ventr_Box4}, Ventrella's Box4 curve on the square grid. 

\noindent$\lr{1,2,1,-2,\mathbf{-1},-2,1,2,1,2,-1,2,1,-2,1,2,-1,2,1,2,\ldots},$\\ \indent c.f.~\ref{sub:ArndtPeano}, Arndt's Peano curve on the square grid.

\noindent$\lr{1,2,\mathbf{-1},-1,-2,-1,2,2,2,1,-2,1,2,1,-2,1,2,1,-2,-2,\ldots},$\\ \indent cf.~\ref{sub:betaOmeg}, $\beta$-$\Omega$ curve on the square grid.

\noindent$\lr{1,2,-1,\mathbf{2},2,1,-2,1,2,1,-2,-2,-1,-2,1,1,2,1,-2,1,\ldots},$\\ \indent c.f.~\ref{sub:HlbrtOrig}, Hilbert's Original on the square grid.

\noindent$\lr{1,2,-1,\mathbf{3},1,-2,-1,4,1,2,-1,-3,1,-2,-1,5,1,2,-1,3,\ldots},$\\ \indent c.f.~\ref{sub:Garycrv} Gray curve on the infinite cubic grid.
	
\noindent$\lr{1,2,\mathbf{3},2,1,4,-3,-2,-1,-2,-3,4,1,2,3,2,1,2,3,-4,\ldots},$\\ \indent c.f.~\ref{sub:ArndtPeano_tsg}, Arndt's Peano curve on truncated square grid.

\noindent$\lr{1,2,3,\mathbf{4},-1,2,3,4,-2,1,-3,4,-1,-2,-3,4,-1,2,3,4,\ldots},$\\ \indent c.f.~\ref{sub:Ventr_V1drgn_8rt}, Ventrella's V1 Dragon curve on the $4$-axes grid.

\noindent$\lr{1,2,3,4,\mathbf{2},1,-4,-3,2,1,-4,-2,-1,-2,-3,1,2,1,-4,-3,1,2,3,4,1,2,3,-1,-2,-1,4,2,2,\ldots},$\\
\indent c.f.~\ref{sub:HlbrtOrig2}, the $4$-axes Hilbert sequence.

\noindent$\lr{1,2,3,4,\mathbf{-2},-1,5,6,6,1,-5,-3,2,1,-5,-3,2,1,-5,-6,-1,-2,-3,-4,-4,2,3,5,1,\ldots},$\\
\indent cf.~\ref{sub:betaOmeg2}, $\beta$-$\Omega$ curve on a $6$-axes grid.

\subsection{Dekking's Gosper-type curve}\label{sub:Dkkng Flwsnk}
	\textbf{sequence:} 
	$\lr{1,1,2,-1,2,1,2,-1,-1,2,1,1,1,2,1,-2,-2,-1,-2,-2,1,2,1,-2,-2,1,\ldots}$
	\vskip 1mm \noindent
	\textbf{in \href{https://oeis.org}{OEIS}:} \seqnum{A356112}
	\vskip 1mm \noindent
	\textbf{alphabet: } $\Sigma_2=\{\pm 1,\pm 2\}$
	\vskip 1mm \noindent
	\textbf{start sequence: } $\lr{1}$
	\vskip 1mm \noindent
	\textbf{substitution: }\\
		$T(\iota)\!=\!(\iota,\iota,\mu\tau_y,-\tau_y,\mu,\iota,\mu\tau_y,-\tau_y,-\iota,\mu\tau_y,\iota,\iota,\tau_y,\mu,\tau_y,-\mu,-\mu,-\tau_y,-\mu,-\mu\tau_y,\tau_y,\mu,\iota,-\mu\tau_y,-\mu\tau_y),$\\
	\indent where $\tau_y=-\tau_x=[1,-2]$ and, as usual, $\mu=[2,-1]$.
	\vskip 1mm \noindent
	\textbf{grid: } The square, plane grid.
	\vskip 1mm \noindent
	\textbf{generators: } $\begin{pmatrix}
		1 & 0 \\ 0 & 1 \\
	\end{pmatrix}$
	\vskip 1mm \noindent

\begin{figure}[H]
	\begin{center}
		\includegraphics[scale=0.31]{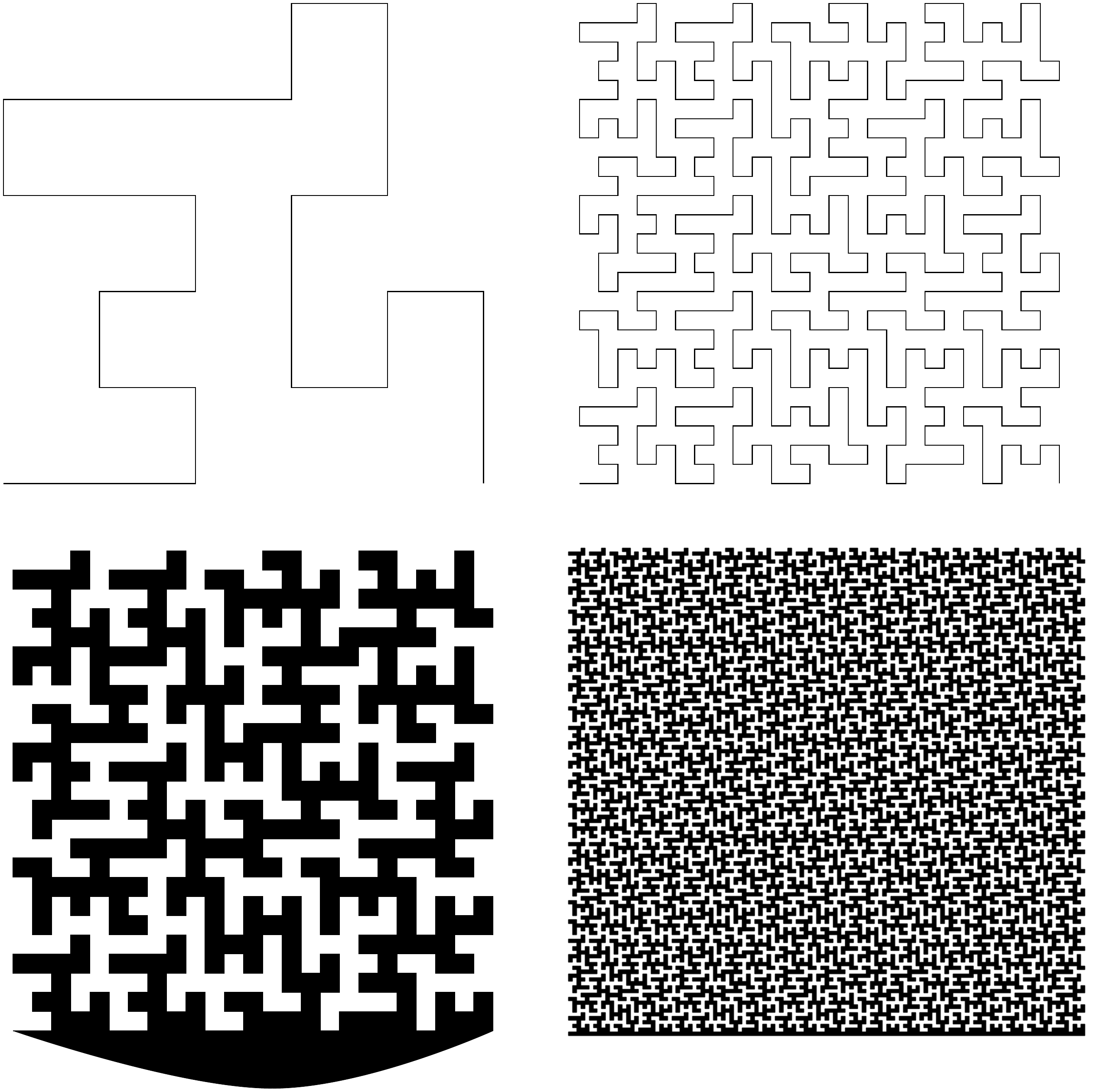}
		\caption{\small{Drawings of the $1^{\text{st}}$ and $2^{\text{nd}}$ approximants in the first row, below the second and third approximants, separating two parts of space in black and white, both with tree-like structures.}}
		\label{fig:Dekking_flowsnake}
	\end{center}	
\end{figure}

\subsection{Dekking's Gosper-type curve on \texorpdfstring{$4$}{4}-axes} \label{sub:Dkkng Flwsnk2}
\begin{description}
	\setlength{\itemsep}{0ex}
\item[sequence:]
$\lr{1,1,2,3,4,1,2,3,-1,2,1,1,-3,4,-3,-4,-4,3,-4,-2,-3,4,1,-2,-2,1,1,2,3,\ldots}$
\item[in \href{https://oeis.org}{OEIS}:] Not (12-12-22)
\item[alphabet:] $\Sigma_4=\{\pm 1,\pm 2, \pm 3, \pm 4\}$
\item[start sequence:] $\lr{1}$
\item[substitution:]
$T(\iota)\!=\!(\iota,\iota,\mu\tau_y,-\tau_y,\mu,\iota,\mu\tau_y,-\tau_y,-\iota,\mu\tau_y,\iota,\iota,\tau_y,\mu,\tau_y,-\mu,-\mu,-\tau_y,-\mu,-\mu\tau_y,\\ \tau_y,\mu,\iota,-\mu\tau_y,-\mu\tau_y),$
 where $\tau_y=[-3,-4,-1,-2]$ and $\mu=[4,3,-2,-1]$.
\item[grid:] The $8^{\text{th}}$-root grid.
\item[generators:] 
$\begin{pmatrix}
	1 & 0 & -\frac{1}{2}\sqrt{2} & \frac{1}{2}\sqrt{2}\\ 0 & 1 & \frac{1}{2}\sqrt{2} & \frac{1}{2}\sqrt{2} \\
\end{pmatrix}$.

\end{description}
\begin{figure}[H]
	\begin{center}
		\includegraphics[scale=0.3]{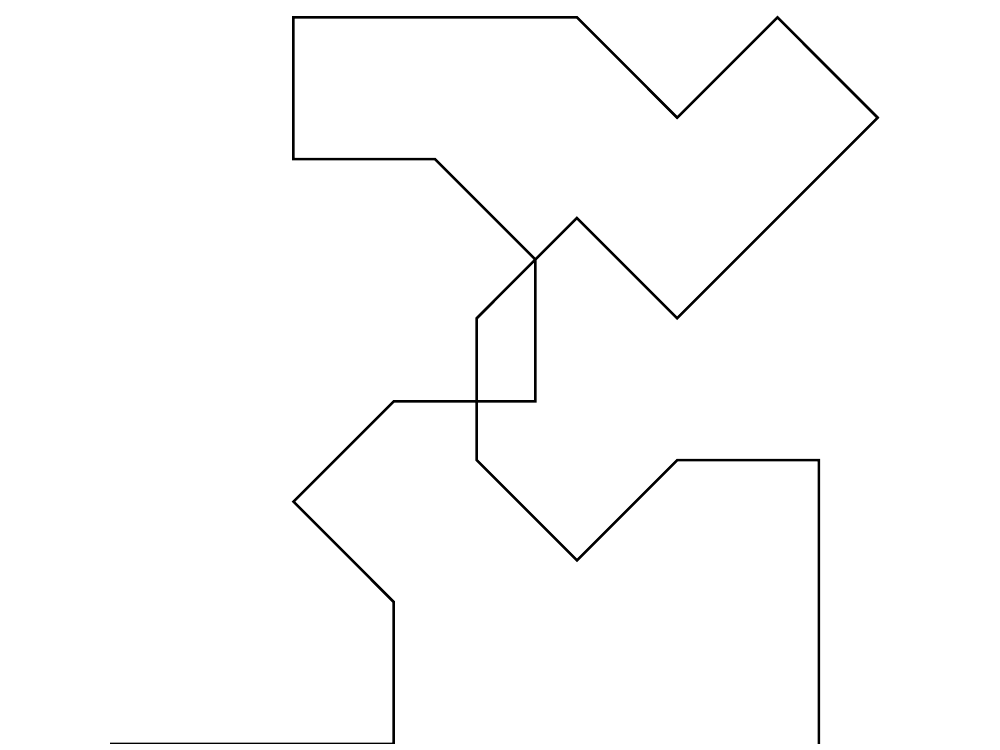}
		\includegraphics[scale=0.4]{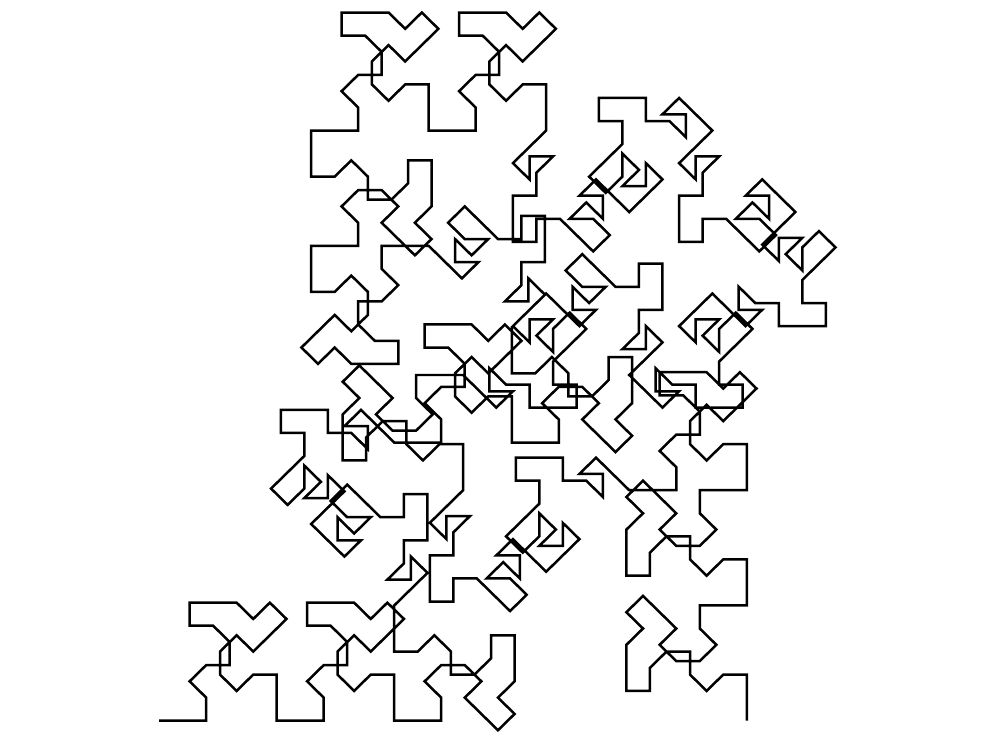}
		\caption{\small{The first $2$ approximants of Dekking's Gosper-type curve on $4$-axes.}}
		\label{fig:Dekking_flowsnake2}
	\end{center}	
\end{figure}

\subsection{Ventrella's Box4}\label{sub:Ventr_Box4}
\begin{description}
	\setlength{\itemsep}{0ex}
	\item[sequence:] $\lr{1,2,1,-2,1,-2,-1,-2,1,-2,1,2,1,2,-1,2,1,-2,1,2,1,2,-1,2,-1,-2,\ldots}$
	\item[in \href{https://oeis.org}{OEIS}:] Not (18-01-2022)
	\item[alphabet:] $\Sigma_2=\{\pm 1,\pm 2\}$
	\item[start sequence:] $\lr{1}$
	\item[substitution:] $T(\iota)=\big(\iota, (-\iota)^k\mu\tau_y\Rev, \tau_y, (-\iota)^{k+1}\mu\Rev\big)$, where $\tau_y=[1,-2]$ and $\mu=[2,-1]$
	\item[grid:] The square grid.
	\item[generators:] $\begin{pmatrix}
		1 & 0 \\ 0 & 1 \\
	\end{pmatrix}$
\end{description}

\begin{figure}[H]
	\includegraphics[scale=0.95]{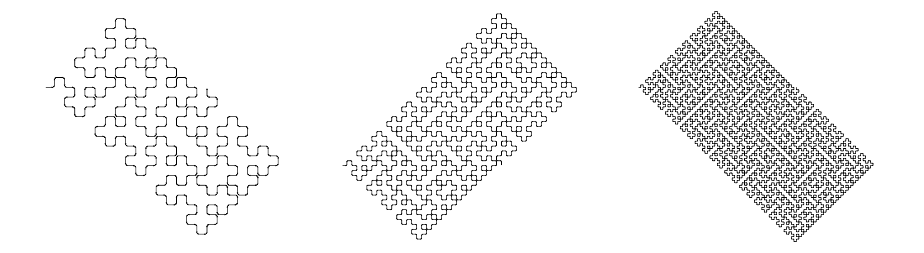}
	\caption{\small The $k$-curves for $k=1,2,3,5$, with corners rounded for the last two.}
\end{figure}

\subsection{Arndt's Peano curve}\label{sub:ArndtPeano}
\begin{description}
	\setlength{\itemsep}{0ex}
	\item[sequence:] $\lr{1,2,1,-2,-1,-2,1,2,1,2,-1,2,1,-2,1,2,-1,2,1,2,1,-2,-1,-2,1,2,1,\ldots}$
	\item[in \href{https://oeis.org}{OEIS}:] Not (18-01-22)
	\item[alphabet:] $\Sigma_2=\{\pm 1,\pm 2\}$
	\item[start sequence:] $\lr{1,2,1,-2,-1,-2,1,2,1}$
	\item[substitution:] $T(\iota)=(\iota,\mu,\iota,\mu^{-1},-,\mu^{-1},\iota,\mu,\iota)$, where $\mu=[2,-1]$.
	\item[grid:] The square grid.
	\item[generators:] $\begin{pmatrix}
		1 & 0 \\ 0 & 1 \\
	\end{pmatrix}$
	\item[simple substitution:]
	\begin{equation*}
		T=\begin{cases}
			1 &\to\lr{ 1,2,1,-2,-1,-2,1,2,1}\\
			2 &\to\lr{ 2,-1,2,1,-2,1,2,-1,2} \\
		\end{cases},
		\text{ or even simpler: }
		T_D=\begin{cases}
			1 &\to\lr{ 1,2,1}\\
			2 &\to\lr{-2,-1,-2} \\
		\end{cases}
	\end{equation*}
	Al this with $T(-x)=-T(x)$.
\end{description}

\begin{figure}[H]
	\centering
	\includegraphics[scale=0.8]{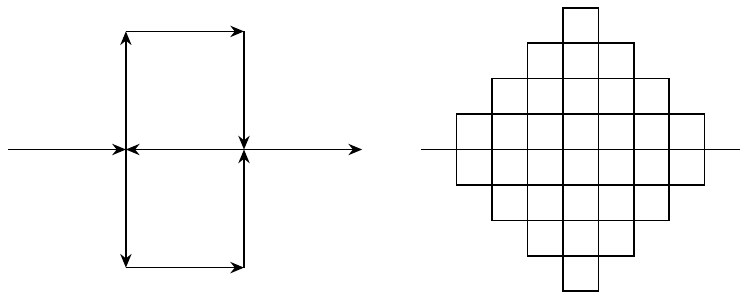}
	\caption{\small The first two approximants of Arndt's R9-1, the Peano curve, on the square grid.} 
	\label{fig:R9-1b}
\end{figure}

\subsection{The \texorpdfstring{$\beta,\Omega$}{beta, Omega} curve}\label{sub:betaOmeg}
\begin{description}
	\setlength{\itemsep}{0ex}
	\item[sequence:] 		$\lr{1,2,-1,-1,-2,-1,2,2,2,1,-2,1,2,1,-2,1,2,1,-2,-2,-1,-2,1,1,1,2,\ldots}$
	\item[in \href{https://oeis.org}{OEIS}:] Not (18-01-2022)
	\item[alphabet:] $\Sigma_2=\{\pm 1,\pm 2\}$
	\item[start sequences:] $\beta_1=\lr{1,2,-1,-1}$;$\beta'_1=\lr{1,-2,-1,-2}$; $\Omega_1=\lr{1,2,-1,2}$
	\item[substitution:] 
	\begin{align*}
		\beta_{k+1}&=\Big(\tau_x(\beta_k),-\mu(\beta_k),\tau_x\mu(\beta'_k),\mu(\Omega_k)\Big) \\
		\beta'_{k+1}&=\Big(\tau_x\mu(\Omega_k),\tau_x\mu(\beta_k),-\mu(\beta'_k),\tau_x(\beta'_k)\Big)\\
		\Omega_{k+1}&=\Big(\tau_x(\beta_k), -\mu(\beta_k),\tau_x\mu(\beta'_k), -\iota(\beta'_k)\Big)
	\end{align*}
	for $k\ge 1$, where $\mu=[2,-1]$ is the minimal grid rotation and $\tau_x=[-1,2]$ the horizontal reflection. Note that $\lim\limits_{k\to\infty} \beta_k=\lim\limits_{k\to\infty}\Omega_k$.
	The normalized curve we obtain by $T_{\beta}(k+1)=\tau_x^k (\beta_{k+1})$.
	\item[grid:] The square grid.
	\item[generators:] $\begin{pmatrix}
		1 & 0 \\ 0 & 1 \\
	\end{pmatrix}$
\end{description}

\begin{figure}[H]
	\centering
	\includegraphics[scale=0.5]{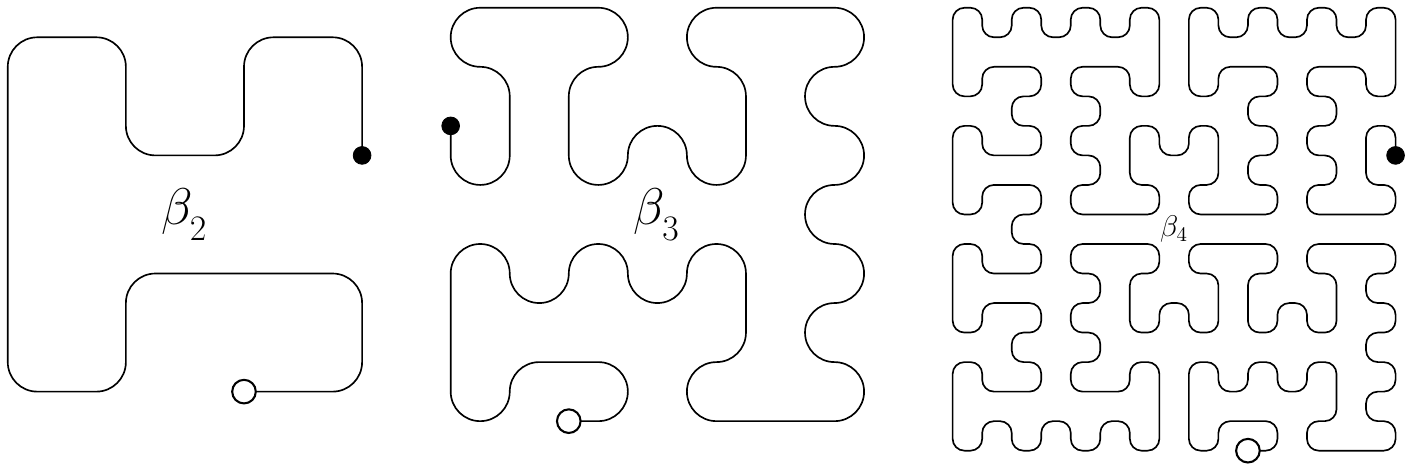}
	\caption{\small The $2,3$- and $4$-curves of the $\beta$ fractal, respectively, with entries $(\circ)$ and exits $(\bullet)$}
	\label{fig:betom23a} 
\end{figure}

\subsection{Hilbert's original and the \texorpdfstring{$4$}{4}-axes Hilbert sequence}\label{sub:HlbrtOrig}
\begin{description}
	\setlength{\itemsep}{0ex}
	\item[sequences:] 		$\lr{1,2,-1,2,2,1,-2,1,2,1,-2,-2,-1,-2,1,1,2,1,-2,1,1,2,-1,2,1,2,-1,\ldots}$,\\
	$\lr{1,2,3,4,2,1,-4,-3,2,1,-4,-2,-1,-2,-3,1,2,1,-4,-3,1,2,3,4,1,2,3,-1,-2,\ldots}.$
	\item[in \href{https://oeis.org}{OEIS}:]\seqnum{A163540}, with $1,2,-1,-2$ replaced by $0,1,2,3$, respectively.
	\item[alphabet:] $\Sigma_2=\{\pm 1,\pm 2\}$
	\item[start sequence:] $\lr{1}$
	\item[substitution:] $A(k+1) = T\big(A(k)\big)=\Big(H(k),\tau_{d}\big(H(k)\big),\tau_{d}\varphi\big(H(k)\big),-\varphi\big(A(k)\big)\Big)$ with \\$A(k)\in\{H(k),\varphi\big(H(k)\big)\}$ and\\
	$\varphi(S)=\varphi\Lr{s(1),s(2),\ldots,s(2^n)}=\Lr{s(1),s(2),\ldots,\tau_d\big(s(2^n)\big)}$ with $\tau_d=[2,1]$.
	\item[grid:] The square grid.
	\item[generators:] $\begin{pmatrix}
		1 & 0 \\ 0 & 1 \\
	\end{pmatrix}$
\end{description}

\begin{figure}[H]
	\centering
	\includegraphics[scale=0.8]{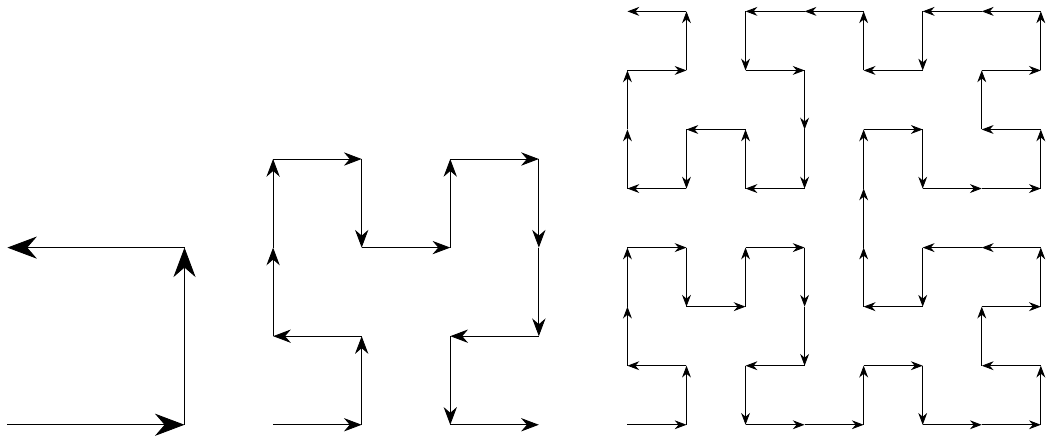}
	\caption{\small The $1$-, $2$- and $3$-curves of the original Hilbert curve.}
	\label{fig:HlbrtOrig}
\end{figure}

\subsection{Gray curve}\label{sub:Garycrv}
\begin{description}
	\setlength{\itemsep}{0ex}
	\item[sequence:]
	$\lr{1,2,-1,3,1,-2,-1,4,1,2,-1,-3,1,-2,-1,5,1,2,-1,3,1,-2,-1,-4,1,\ldots}$
	\item[in \href{https://oeis.org}{OEIS}:] \href{https://oeis.org/A164677}{A164677} (02-11-2021)
	\item[alphabet:] $\Sigma_{\N}=\{1,2,3,\ldots\}=\N\setminus\{0\}$
	\item[start sequence:] $\lr{1}$
	\item[substitution:] 
	\begin{equation*}
		\begin{cases}
			T(x)=\Lr{1,x+\sgn(x)}\text{ for } |x|=1,\\
			T(x)=\Lr{-1,x+\sgn(x)}\text{ for } |x|\not=1 \\
		\end{cases}
	\end{equation*}
	\item[grid:] The cubic grid $[0,1]^{\N}$
	\item[generators:] $\big\{(\delta_{1k},\delta_{2k},\delta_{3k},\ldots)\big\}; k=1,2,\ldots$, where $\delta_{jk}$ is Kronecker delta.
\end{description}

\begin{figure}[H]
	\centering
	\includegraphics[scale=0.8]{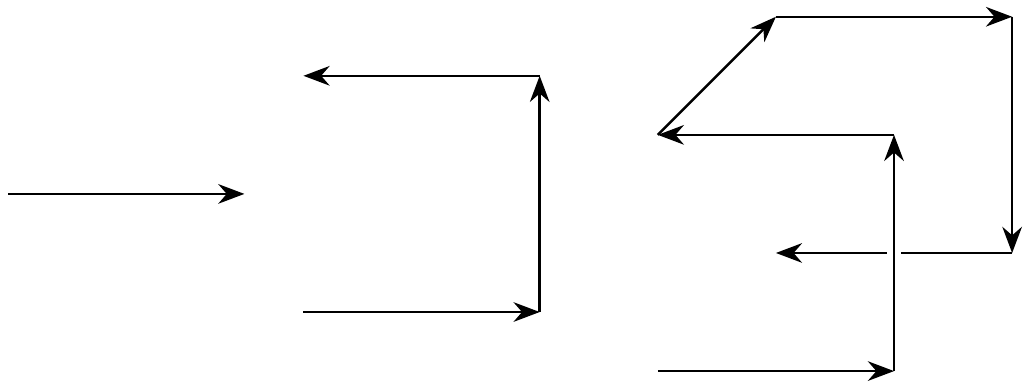}
	\caption{\small The first three approximants of the Gray curve.} 
	\label{fig:graycrv123a}
\end{figure}

\subsection{Arndt's Peano curve on the truncated square grid} \label{sub:ArndtPeano_tsg}
\begin{description}
	\setlength{\itemsep}{0ex}
	\item[sequence:]
	$\lr{1,2,3,2,1,4,-3,-2,-1,-2,-3,4,1,2,3,2,1,2,3,-4,-1,-4,3,2,1,4,-3,\ldots}$
	\item[in \href{https://oeis.org}{OEIS}:] Not (18-01-2022)
	\item[alphabet:] $\Sigma_4=\{\pm 1,\pm 2,\pm 3,\pm 4\}$
	\item[start sequence:] subsequent pairs of \ref{sub:ArndtPeano}, \nameref{sub:ArndtPeano}
	\item[substitution:] 
	\begin{equation*}
		T=\begin{cases}
			\lr{1,2}&\to\;\lr{1,2}\\
			\lr{1,-2}& \to\;\lr{1,4}\\
			\lr{2,1}&\to\;\lr{3,2} \\
			\lr{2,-1}& \to\;\lr{3,-4} \\
		\end{cases}
		\quad\text{ together with }T\lr{-x,-y} = -T\lr{x,y}
	\end{equation*}
	\item[grid:] The truncated square grid spanned by the $8^\text{th}$-roots of unity (c.f.~\cref{fig:sqrdiag8roots}).
	\item[generators:] 
	$\begin{pmatrix}
		1 &  \frac{1}{2}\sqrt{2} &0 & -\frac{1}{2}\sqrt{2}\\ 0 & \frac{1}{2}\sqrt{2} & 1 & \frac{1}{2}\sqrt{2} \\
	\end{pmatrix}$.
\end{description}

\begin{figure}[H]
	\centering
	\includegraphics[scale=1]{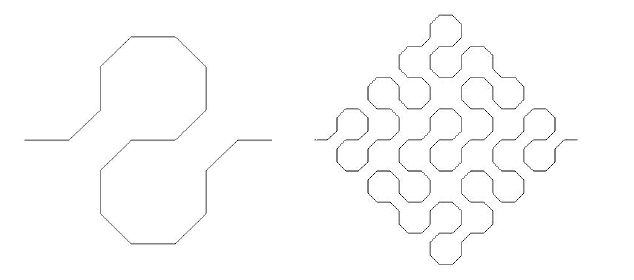}
	\caption{\small The first two approximants of Arndt's Peano on the truncated square grid.} 
	\label{fig:TSqr_R9-1b}
\end{figure}

\subsection{Ventrella's V1 Dragon on \texorpdfstring{$8^{\text{th}}$}{8-th}-root plane grid}\label{sub:Ventr_V1drgn_8rt}
\begin{description}
	\setlength{\itemsep}{0ex}
	\item[sequence:] 		$\lr{1,2,3,4,-1,2,3,4,-2,1,-3,4,-1,-2,-3,4,-1,2,3,4,-2,1,-3,4,-2,1,\ldots}$
	\item[in \href{https://oeis.org}{OEIS}:] Not (18-01-2022)
	\item[alphabet:] $\Sigma_4=\{\pm 1,\pm 2, \pm 3, \pm 4\}$
	\item[start sequence:]$\lr{1}$
	\item[substitution:] $T(\iota)=\big(\iota, \Rev\mu\,^2, \mu\,^3 \big)$, where $\mu=[-4,3,-1,-2]$
	\item[grid:] The $8^{\text{th}}$-root plane grid.
	\item[generators:] 
	$\begin{pmatrix}
		1 & 0 & -\frac{1}{2}\sqrt{2} & -\frac{1}{2}\sqrt{2}\\ 0 & 1 & \frac{1}{2}\sqrt{2} & -\frac{1}{2}\sqrt{2} \\
	\end{pmatrix}$.
\end{description}

\begin{figure}[H]
	\centering
	\includegraphics[scale=0.8]{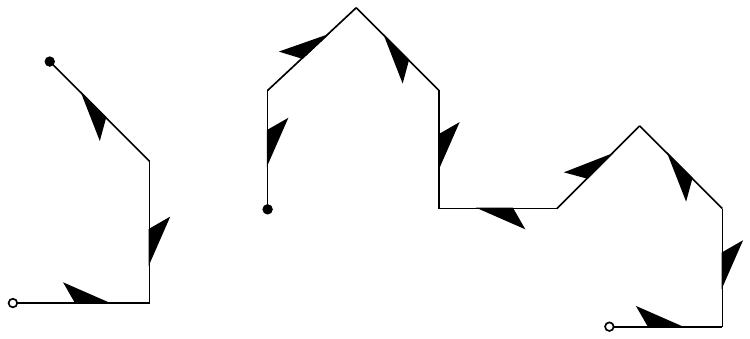}
	\caption{\small The $1$- and $2$-curves on the $8^{\text{th}}$-root grid.}
	\label{fig:V1drgnrt8a}
\end{figure}

\noindent Below the same fractal sequence as above, but on the square-diagonal grid, together with its \emph{length sequence} $_lS$.
\begin{description}
	\setlength{\itemsep}{0ex}
	\item[length sequence:] $\lr{1,1,\sqrt{2},\sqrt{2},1,1,\sqrt{2},\sqrt{2},2,2,\sqrt{2},\sqrt{2},1,1,\sqrt{2},\sqrt{2},1,1,\sqrt{2},\sqrt{2},2,2\ldots}$
	\item[length substitution:] $_lT(\iota)=\big(\iota, \Rev, \iota*\sqrt{2} \big)$
	\item[in \href{https://oeis.org}{OEIS}:] 
	$\log_{\sqrt{2}}\left(_lS\right)=\lr{0,0,1,1,0,0,1,1,2,2,1,1,0,0,1,1,0,0,1,1,2,2,1,1,2,2,3,3,2,\ldots}$
	that is the double \big(i.e., $\lr{x,x}$ versus $\lr{x}$\big) of $\mathbf{A062756}$
\end{description}

\begin{figure}[H]
	\centering
	\includegraphics[scale=0.7]{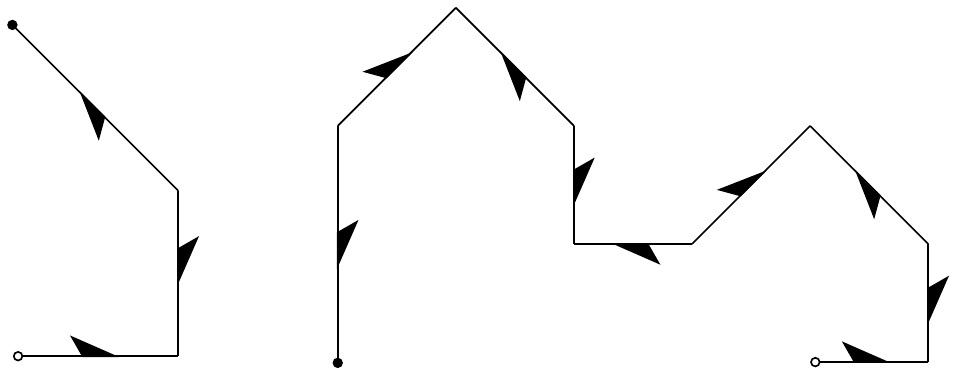}
	\caption{\small The $1$- and $2$-curves of Ventrella's V1 Dragon on the square-diagonal grid.}
	\label{fig:V1drgn_a}
\end{figure}

\subsection{Hilbert's original on the \texorpdfstring{$4$}{4}-axes grid}\label{sub:HlbrtOrig2}
\begin{description}
	\setlength{\itemsep}{0ex}
	\item[sequences:] 			$\lr{1,2,3,4,2,1,-4,-3,2,1,-4,-2,-1,-2,-3,1,2,1,-4,-3,1,2,3,4,1,2,3,\ldots}.$
	\item[in \href{https://oeis.org}{OEIS}:] Not (12-12-2022)
	\item[alphabet:] $\Sigma_4=\{\pm 1,\pm 2,\pm 3, \pm 4\}$
	\item[start sequence:] $\lr{1}$
	\item[substitution:]
	\begin{equation*}
		T=\begin{cases}
			1 &\to \lr{ 1,2,3,4}\\
			2 &\to \lr{ 2,1,-4,-3} \\
			3 &\to \lr{ 2,1,-4,-2} \\
			4 &\to \lr{ -1,-2,-3,1}\\
		\end{cases}
		\quad\; T(-x)=-T(x)\text{ for }x\in \{1, 2, 3, 4\}.
	\end{equation*}
	\item[grid:] The $8^{\text{th}}$-root grid.
	\item[generators:] 
	$\begin{pmatrix}
		1 & 0 & -\frac{1}{2}\sqrt{2} & -\frac{1}{2}\sqrt{2}\\ 0 & 1 & -\frac{1}{2}\sqrt{2} & \frac{1}{2}\sqrt{2} \\
	\end{pmatrix}$.
\end{description}
After replacing $\mp 3$ by $\pm 1$ and $\pm 4$ by $\pm 2$, we get our Hilbert sequence, \cref{sub:HlbrtOrig}.

\begin{figure}[H]
	\centering
	\includegraphics[scale=0.50]{Hilbert_4_s45.pdf}\quad\quad
	\includegraphics[trim=0 -0.8cm 0 0, scale=1]{8th_roots_grid_c.pdf}
	\caption{\small The $4$-axes Hilbert fractal on the renumbered $8^\text{th}$-root grid.} 
	\label{fig:hilbert4ext1}
\end{figure}

\subsection{The \texorpdfstring{$\beta,\Omega$}{beta, Omega} curve on \texorpdfstring{$6$}{6} axes}\label{sub:betaOmeg2}
\begin{description}
	\setlength{\itemsep}{0ex}
	\item[sequence:]$\lr{1,2,3,4,-2,-1,5,6,6,1,-5,-3,2,1,-5,-3,2,1,-5,-6,-1,-2,-3,-4,-4,2,\ldots}$
	\item[in \href{https://oeis.org}{OEIS}:] Not (12-12-2022)
	\item[alphabet:] $\Sigma_6=\{\pm 1,\pm 2,\ldots,\pm 6\}$
	\item[start sequences:] $\lr{1}$
	\item[substitution:] With $T(-x)=-T(x)$ we get
	\begin{equation*}
		T=\begin{cases}
			1 &\to\lr{ 1, 6,-2,-3}\\
			2 &\to\lr{ -5,-1, 4,-2}\\
			3 &\to\lr{ -6,-1, 4,-2} \\
			4 &\to\lr{ 3, 6,-2, 4}\\
			5 &\to\lr{ 1, 6,-2, 4} \\
			6 &\to\lr{ -6,-1, 4, 5} \\
		\end{cases}
	\end{equation*}
	To obtain de $\beta,\Omega$ sequence (\cref{sub:betaOmeg}), we replace $\pm 1,\pm 2,\pm 3$ by axis $\{\pm 1\}$ and $\pm 4,\pm 5,\pm 6$ by axis $\{\pm2\}$.
	\item[grid:] The $12^{\text{th}}$-root grid.
	\item[generators:] 
	$\begin{pmatrix}
		1 & \frac{1}{2}\sqrt{3} & -\frac{1}{2} & -\frac{1}{2}\sqrt{3} & \frac{1}{2} & 0\\
		0 & -\frac{1}{2} & -\frac{1}{2}\sqrt{3} & -\frac{1}{2} & -\frac{1}{2}\sqrt{3} & -1\\
	\end{pmatrix}$.
\end{description}
\begin{figure}[H]
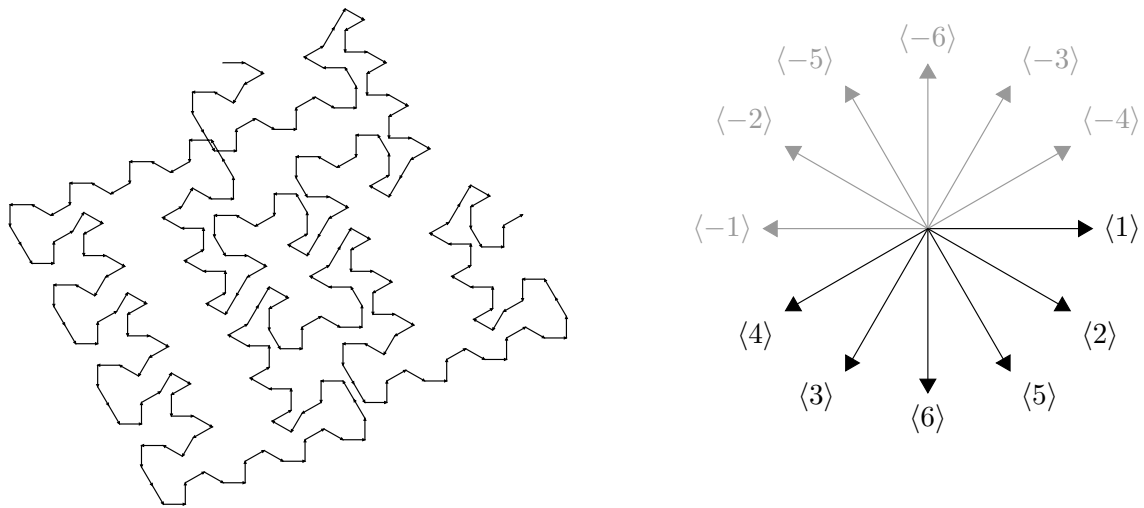

	\centering
	\includegraphics[scale=0.55]{12th_root_bO.pdf}
	\includegraphics[trim=-0.3cm -0.8cm 0 0, scale=1.1]{12th_roots.pdf}
	\caption{\small The six axes fractal on the renumbered $12^\text{th}$-root grid (page \pageref{fig:6dimcrv}).} 
\end{figure}

\bigskip
\hrule
\bigskip

\noindent 2010 \emph{Mathematics Subject Classification}:
Primary 11B85; Secondary 28A80, 05Cxx.

\noindent \emph{Keywords}: fractal sequence, number sequence, signed permutation, approximant, grid, digi\-set, isometry, direction, substitution.

\end{document}